\documentclass[12pt]{article} 
\pdfoutput=1
\usepackage{jheppub}
\usepackage{slashed}
\usepackage[utf8]{inputenc}
\usepackage[fleqn]{mathtools}
\usepackage[fleqn]{amsmath}
\usepackage{amssymb,mathrsfs}
\usepackage{setspace}
\usepackage{graphicx,floatflt,rotate, cancel}
\usepackage{bm}
\usepackage{soul}
\usepackage{color}
\usepackage{cancel}
\usepackage{xcolor}
\usepackage{graphicx}
\usepackage{multirow}
\usepackage{array}
\usepackage{float}
\usepackage[toc,page]{appendix}
\usepackage{diagbox}
\usepackage[mathscr]{euscript}
\usepackage{cleveref}
\usepackage{soul}
\usepackage{pifont}
\usepackage[normalem]{ulem}
\usepackage[numbers]{natbib}
\usepackage{notoccite}
\usepackage{comment}
\makeatletter
\gdef\@fpheader{}
\makeatother

\usepackage[utf8]{inputenc}
\DeclareUnicodeCharacter{03BC}{$\mu$}

\DeclareUnicodeCharacter{2212}{\ensuremath{-}}
\definecolor{orange}{rgb}{1,0.5,0}

\title{Chasing Long-Lived Doubly Charged Scalars at Future Lepton Colliders}
\author[a,b]{Nandini Das,}
\author[c]{Dilip Kumar Ghosh,}
\author[d]{Nivedita Ghosh,}
\author[e]{Ritesh K. Singh}
\affiliation[a]{Department of Physics, SGTB Khalsa College, Delhi 110007, India}
\affiliation[b]{Department of Physics and Astrophysics, University of Delhi, Delhi 110007, India}
\affiliation[c]{School of Physical Sciences, Indian Association for the Cultivation of Science, 2A $\&$ 2B, Raja S.C. Mullick Road, Kolkata 700032, India}
\affiliation[d]{Kavli IPMU (WPI), UTIAS, University of Tokyo,  Kashiwa, Chiba, 277-8583, Japan}
\affiliation[e]{ Department of Physical Sciences, Indian Institute of Science Education and Research Kolkata, Mohanpur, 741246, Nadia, India}

\emailAdd{nandinidas.rs@gmail.com}
\emailAdd{tpdkg@iacs.res.in}
\emailAdd{nivedita.ghosh@ipmu.jp}
\emailAdd{ritesh.singh@iiserkol.ac.in}

\abstract{  
We come up with a novel search strategy for long-lived doubly charged
scalars at future proposed lepton colliders.
The doubly charged scalar studied in this work belongs to an $SU(2)_L$ complex scalar triplet that accounts for tiny neutrino masses via the Type-II Seesaw mechanism. For scalar masses $\lesssim 200 $ GeV
and appropriate values of the triplet vacuum expectation value, this state can be long-lived and decay predominantly into like-sign
muon pairs (e.g. $\mu^+\mu^+ $ or $\mu^-\mu^-$), producing
distinctive displaced-vertex signals. We investigate the pair production of these scalars at the International
Linear Collider (ILC) and a prospective muon collider, considering their
planned center-of-mass energies. Incorporating theoretical and experimental
constraints, we study the resulting signature of four leptons accompanied
by missing transverse energy. Displaced vertices offer direct evidence of
the scalar's long lifetime, while we further show that the invariant mass distribution of same-sign dilepton pairs serves as a powerful complementary probe for discovering doubly charged Higgs bosons
at both the ILC and muon collider.}

\begin{document}

\maketitle

\setcounter{footnote}{0}  
\renewcommand{\thefootnote}{\arabic{footnote}}

\section{Introduction}~\label{intro}
A decade after the Higgs discovery~\cite{ATLAS:2012yve,CMS:2012qbp}, the non-observation of any Beyond the Standard Model (BSM) signals at the collider urges us to revisit the underlying search techniques. Most of the searches at the colliders assume that the new physics (NP) particles decay promptly. But this need not be true, as even within the SM, we have mesons and baryons which are Long-Lived Particles (LLPs)~\cite{ParticleDataGroup:2024cfk}, i.e., they have a proper decay length $c\tau \gtrsim \cal{O}$ (mm) in the detector. The possibility that NP particles can be long-lived has important consequences in the collider searches~\cite{Lee:2018pag,Alimena:2019zri,Knapen:2022afb,Bose:2022obr,Jeanty:2025wai}. In particular, searches for scalar LLPs have long been one of the central focus of beam-dump experiments such as SHiP~\cite{Lee:2019lfo}, NA62~\cite{Lanfranchi:2017wzl}, and NA64~\cite{NA64:2020qwq}, as well as in experiments and phenomenology at the Large Hadron Collider (LHC)~\cite{Belyaev:2015ldo,FASER:2018eoc,CODEX-b:2019jve,Bauer:2019vqk,Cepeda:2021rql,Cerci:2021nlb,MoEDAL-MAPP:2022kyr,Zaffaroni:2022xlu,Curtin:2023skh,Akhmedov:2024rvp,ATLAS:2024itc,Lu:2024ade,ATLAS:2024vnc,Niedziela:2024khw,Alimena:2025kjv,Fang:2025iui}, the International Linear Collider (ILC)~\cite{Klamka:2024gvd}, and proposed future circular colliders (FCC)~\cite{Bhattacherjee:2021rml,Bhattacherjee:2023plj,Bhattacherjee:2025dlu,Bhattacherjee:2025gwo,Bhattacherjee:2025pxg}. The phenomenology of scalar LLPs has also been extensively explored at low-energy $e^{+}e^{-}$ colliders, such as Belle~II~\cite{Filimonova:2019tuy}, as well as at neutrino facilities~\cite{Dev:2021qjj,Batell:2022xau}.
 \\

In a different context, the generation of tiny, nonzero neutrino masses through the
seesaw mechanism can be realized in the Type-II seesaw model~\cite{Schechter:1980gr,Magg:1980ut,Cheng:1980qt,Lazarides:1980nt,Mohapatra:1980yp}. The scenario extends the SM by introducing a $SU(2)_L$ complex triplet scalar, where lepton number is violated by two units through a trilinear coupling $\mu$ in the scalar potential. Neutrinos acquire Majorana masses via the Yukawa interaction between the triplet and the lepton doublets, leading to
$ m_\nu \sim \frac{Y_\Delta\, \mu\, v_d^2}{M_\Delta^2}$,
with $M_\Delta$ being the triplet mass and $v_d$ being the vacuum expectation value of the SM Higgs doublet (\textbf{vev}). The smallness of $m_\nu$ can thus be explained by a naturally small $\mu$ term, protected by symmetry in the sense of \textit{’t Hooft} naturalness~\cite{tHooft:1979rat}. The same $\mu$ term generates the triplet vev
$v_t \sim \frac{\mu\, v_d^2}{M_\Delta^2}$, which is induced after the electroweak symmetry breaking. An appropriate choice of $(\mu$ or $v_t)$ and $Y_\Delta$, consistent with current neutrino oscillation data~\cite{deSalas:2017kay}, allows one to obtain tiny neutrino masses even for triplet scalars near the electroweak scale ($M_\Delta \sim v_d$). Additionally, the presence of the exotic singly and doubly charged scalars makes this model phenomenologically appealing for collider studies. In particular, the most striking signature of the triplet extension is the appearance of a doubly charged scalar, which offers distinctive and experimentally accessible collider signals. \\

Depending on the triplet vev, the doubly charged scalar can decay to a pair of dileptons of the same-sign ($v_t < 10^{-4} ~\rm GeV$), or a pair of gauge bosons $(v_t > 10^{-4} ~\rm GeV$)~\cite{FileviezPerez:2008jbu,Melfo:2011nx,Aoki:2011pz}.  The ATLAS \cite{ATLAS:2018ceg,ATLAS:2021jol,ATLAS:2022pbd,ATLAS:2022yzd,ATLAS:2024itc} and CMS \cite{CMS:2017pet,CMS:2017fhs,CMS:2021wlt} collaborations have searched for these doubly charged scalars $H^{\pm\pm}$ in both pure leptonic and $W^{\pm}W^{\pm}$ final states at $\sqrt{s}=$ 13 TeV, assuming prompt decay of the scalar. The non-observance of any significant excess at the LHC has put an upper limit on the mass of the doubly charged scalar to be $> 1080$ GeV for $v_t < 10^{-4} ~\rm GeV$~\cite{ATLAS:2022pbd}. However, one should keep in mind that all these null results are derived assuming that the doubly-charged scalar decays promptly. The possibility that these scalar particles may not decay promptly, but instead exhibit long-lived behavior, carries important implications for their search strategies at the collider. Searches for such heavy stable charged particles (HSCPs) have already been performed by ATLAS~\cite{ATLAS:2022cob,ATLAS:2023zxo} and CMS~\cite{CMS:2016kce,CMS:2025vwv}. If the decays of a long-lived particle are non-prompt but occur inside the detector volume, displaced secondary vertices can also serve as a powerful probe. This possibility has recently been explored in recent literature in the context of the Type-II seesaw model~\cite{BhupalDev:2018tox,Antusch:2018svb}. However, the search for displaced vertex search for the doubly-charged scalars at the lepton colliders is scarce in the literature. \\

In this paper, we rekindle the idea of probing the doubly-charged scalar as a long-lived particle in future precision collider facilities such as the International Linear Collider (ILC)~\cite{ILC:2007oiw,ILC:2007bjz,ILC:2007vrf,Evans:2015xpt,ILC:2019gyn,Abe:2025yur} and the International Muon Collider Collaboration (IMCC) ~\cite{IMCC,Delahaye:2019omf,Schulte:2019bdl,Schulte:2022brl,InternationalMuonCollider:2024jyv,Begel:2025ldu,InternationalMuonCollider:2025sys}. Compared to the hadron colliders, because of the added advantage of being less noisy, the lepton colliders may act as a discovery machine for the doubly charged Higgs boson. To be more specific, we produce the doubly charged scalar in pairs in the proposed 500 GeV ILC machine with $\tt{4~ab^{-1}}$ integrated luminosity~\cite{ILCInternationalDevelopmentTeam:2022izu} whereas we pair produce the doubly charged scalars with a pair of fermions in the proposed 10 TeV muon collider with $\tt{10~ab^{-1}}$ luminosity~\cite{Costantini:2020stv,Han:2020uid}. We find out that owing to the cleaner environment of the colliders, with a few to a few hundred $\rm fb^{-1}$ of luminosity, we can discover the doubly charged scalar in the mass range $\in[100:180]$ GeV by looking at the long-lived di-muonic decays of the scalar. This mass range, as discussed in~\cite{Antusch:2018svb}, corresponds to a region that remains difficult to probe at the LHC and therefore represents an LHC blind spot. Our analysis demonstrates that future $e^+e^-$ and $\mu^+\mu^-$ colliders can effectively cover this blind-spot region.\\

The paper is organized as follows. In Section~\ref{model}, we briefly discuss the model, followed by theoretical and experimental constraints on our model parameter space in Section~\ref{cons}. We motivate our analysis strategy and results in Section~\ref{result}. Finally, we conclude in Section~\ref{conc}.

\section{Model}~\label{model}

In this section, we briefly outline the Type-II seesaw model, which augments the Standard Model (SM) with a $SU(2)_L$ complex triplet scalar field $\Delta$ of hypercharge $Y=2$ given by
\begin{eqnarray}
\Delta &=& \frac{\sigma^i}{\sqrt 2}\Delta_i = \left(\begin{array}{cc}
\delta^+/\sqrt 2 & \delta^{++}\\
\delta^0 & -\delta^+/\sqrt 2
\end{array}
\right),
\end{eqnarray}
\noindent Here $\sigma_i$'s correspond to the Pauli matrices where $i$ runs from one to three and  $\Delta_1=(\delta^{++}+\delta^0)/\sqrt 2,~\Delta_2=i(\delta^{++}-\delta^0)/\sqrt 2,~\Delta_3=\delta^+$.
The Lagrangian of the model, incorporating the SM with the triplet sector, is given by
\begin{eqnarray}
{\cal L} = {\cal L}_{\rm Yukawa}^{\Delta +{\rm SM}}+ \mathcal{L}_{\rm Scalar}^{\Delta+ H}+\mathcal{L}^{\rm SM}_{\rm Gauge}
\label{lag}
\end{eqnarray}
where the third term consists of the gauge boson Lagrangian of the SM and the gauged kinetic terms for the fermions. The scalar part is composed of
\begin{equation}
   \mathcal{L}_{\rm Scalar}^{\Delta+ H}= {\cal L}_{\rm Kinetic}-V(\Phi,\Delta),
\end{equation}
The kinetic and Yukawa interactions are, respectively, \cite{Arhrib:2011uy}
\begin{eqnarray}
\label{y2kinetic}
{\cal L}_{\rm Kinetic} &=&\left(D_\mu\Phi\right)^\dag \left(D^\mu\Phi\right) +{\rm Tr}\left[\left(D_\mu \Delta\right)^\dag \left(D^\mu\Delta\right)\right], \\
\label{y2Yukawa}
{\cal L}^{\rm Yukawa}_{\Delta+{\rm SM}} &=& {\cal L}_{\rm Yukawa}^{\rm SM}- \left(Y_\Delta\right)_{ij} L_i^{\sf T}Ci\sigma_2\Delta L_j+{\rm h.c.}\, .
\end{eqnarray}

Here, $\Phi^{\sf T}=(\phi^+ ~~~\phi^0)$ is the SM scalar doublet, $L$ represents 
the $SU(2)_L$ left-handed lepton doublet, $Y_\Delta$ is the Yukawa coupling, and $C$ is the Dirac charge conjugation matrix. 
The covariant  derivative of the scalar triplet field is given by:
\begin{equation} 
D_\mu \Delta = \partial_\mu \Delta + i\frac{g}{2}[\sigma^a W_\mu^a,\Delta]
+i g^\prime B_\mu \Delta \qquad (a=1,2,3).
\end{equation}
$\sigma^a$ are the Pauli matrices, $g$ and $g^\prime$ are the gauge coupling constants of the $SU(2)_L$ and $U(1)_Y$ group, respectively.

The most general scalar potential can be written as \cite{Arhrib:2011uy}:
\begin{eqnarray} 
V(\Phi,\Delta) &=& -m^2_\Phi(\Phi^\dag \Phi)+\frac{\lambda }{4}(\Phi^\dag \Phi)^2+M^2_\Delta {\rm Tr}(\Delta ^\dag \Delta)+ \left(\mu \Phi^{\sf T}i\sigma_2\Delta^\dag\Phi+{\rm h.c.}\right)+\,\nonumber\\
&& \lambda _1(\Phi^\dag\Phi){\rm Tr}(\Delta ^\dag \Delta)+\lambda _2\left[{\rm Tr}(\Delta ^\dag \Delta)\right]^2+\lambda _3{\rm Tr}(\Delta ^\dag \Delta)^2+\,\nonumber\\
&&\lambda _4\Phi^\dag\Delta\Delta^\dag\Phi.
\label{eq:Vpd}
\end{eqnarray}

In general, all the parameters of the potential can be complex. However, for simplicity, we have assumed that they are real.

After EWSB, the neutral component of the scalar fields expanded around the respective vevs can be parametrized as 
\begin{eqnarray}
\Phi &=& \frac{1}{\sqrt{2}}\left(\begin{array}{c}
\sqrt{2}\chi_d^+ \\
v_d + h_d + i\eta_d \\
\end{array}\right) \qquad 
\Delta = \frac{1}{\sqrt 2}\left(\begin{array}{cc}
\delta^+ & \sqrt{2}\delta^{++}\\
v_t + h_t + i\eta_t & -\delta^+
\end{array}
\right)\;. 
\end{eqnarray}
where $v_d$ and $v_t$ denote the doublet and triplet vacuum expectation values (VEV), respectively.
Similarly, minimizing the potential, the two mass parameters of the potential can be written in terms of the other free parameters as
\begin{equation}
m_\Phi^2 = \lambda \frac{v_d^2}{4} - \sqrt{2} \mu v_t + \frac{(\lambda_1+\lambda_4)}{2}v_t^2 \,,
\label{eq:minim1}
\end{equation}
\begin{equation}
M_\Delta^2 = \frac{\mu v_d^2}{\sqrt{2}v_t} - \frac{\lambda_1 + \lambda_4}{2}v_d^2 - (\lambda_2 + \lambda_3)v_t^2 \,,
\label{eq:minim}
\end{equation}
The scalar spectrum of the model is brought down to the mass basis from the gauge basis by orthogonal rotation in the CP-even, CP-odd, and charged sector
 with rotation angles $\alpha, \beta $ and $\beta^\prime$ 
respectively giving us seven physical Higgs bosons, namely two doubly charged 
$H^{\pm\pm}$, two singly charged $H^\pm$, two CP-even neutral ($h,H$) and a  CP-odd ($A$) scalar.  For more details, see~\cite{Ghosh:2017pxl}. Although no extension of the SM gauge sector is needed for the inclusion of this triplet scalar, the electroweak gauge boson masses ($M_W, M_Z$) get a finite contribution from the triplet VEV $(v_t)$ at tree level, which in turn modifies the $\rho$ parameter. Precision electroweak data constrain the $\rho$ parameter to be very close to its SM value of unity~\cite{ParticleDataGroup:2024cfk}, which imposes an upper bound on $v_t$ of $v_t \lesssim$ 2.1 GeV~\cite{Antusch:2018svb}.
On the other hand, the non-standard triplet Yukawa Lagrangian gives rise to the neutrino mass where the mass matrix is given by $M_\nu = \sqrt{2}{v_t Y_\Delta}$. This mass matrix is diagonalized by the neutrino mixing matrix, i.e., the Pontecorvo–Maki–Nakagawa–Sakata (PMNS) matrix, whose elements are determined from neutrino oscillation data for a given neutrino mass hierarchy~\cite{BhupalDev:2013xol,Bonilla:2015jdf}. Consequently, a neutrino mass of ${\cal O}(0.1)$ eV can be realized by appropriately tuning either the triplet VEV $(v_t)$ or the Yukawa coupling $(Y_\Delta)$. For $v_t \sim {\cal O} (~\rm GeV)$, the Yukawa coupling must be small, whereas an order-one neutrino Yukawa coupling $(Y_\Delta \sim {\cal O} (1))$ requires $v_t < {\cal O}(10^{-2}~\rm eV)$. These two cases represent the extreme limits of the triplet VEV. The choice of the triplet vev plays an important role in the benchmark selections and constraint analysis
discussed in the following section.

\section{Theoretical and Experimental Constraints}~\label{cons}

In this section, we provide a brief overview of the existing theoretical and experimental constraints applicable to our benchmark choices:
\begin{itemize}

\item {\bf {Vacuum stability:}} To have the scalar potential  bounded from below in all directions in field space, we have taken into account all the necessary and sufficient conditions
on the scalar quartic couplings mentioned in~\cite{Arhrib:2011uy}.

\item {\bf Perturbative unitarity:} 
The requirement of perturbativity of the quartic couplings as well as the tree-level unitarity of the 
scattering amplitude for all $2\to2$ processes requires the ${\cal S}$-matrix eigenvalues to be bounded from above ($|a_i| < 16 \pi$). The set of unitarity conditions received from the above requirements for this scalar setup \cite{Arhrib:2011uy} is satisfied by the benchmark points considered in our analysis. 

\item {\bf Constraints from electroweak precision test:}
The presence of the new exotic scalar bosons contributes to the electroweak precision observables, parametrized by the oblique coefficients S, T, and 
U~\cite{Lavoura:1993nq, Aoki:2012jj}. Among these, the strongest
bound comes from the T-parameter, which imposes a strict limit on the 
mass splitting between the doubly and singly charged scalars, 
$\Delta M \equiv \mid m_{H^{\pm \pm}} - m_{H^\pm}\mid \equiv \mid m_{H^{\pm }} - m_{H}/m_{A}\mid$ which
should be $\lesssim 40$ GeV~\cite{Chun:2012jw}. Our scalar spectrum corresponding to the multiplet, being almost degenerate, is safe from this constraint. 

\item {\bf Lepton flavor violation:}
Due to the presence of the Higgs triplet in the model, lepton flavor violation is induced at tree level mediated by the doubly charged scalar. 
The process $\mu \to 3\,e $ puts the stringent constraint with $Br(\mu \to e e e) \lesssim 1.0 \times 10^{-12}$ measured by the SINDRUM experiment \cite{BELLGARDT19881} among all the possible decay channels. The next lepton number violating process $\ell_i \to \ell_j \gamma$ arises at the loop level with contributions from the doubly and singly charged scalar. The stringent constraint among all the channels is given by the MEG collaboration \cite{MEG:2016leq} on the branching ratio of $\mu \to e \gamma$ to be less than $4.2 \times 10^{-13}$. However, the choice of $v_t$ and doubly charged scalar masses we are interested in is safe from these constraints.

\item{\bf {Higgs Signal Strength}:}
The observed SM-like Higgs boson with a mass of 125~GeV, and hence its decay widths, must remain consistent with the Higgs measurements reported at the LHC. In particular, the total width can receive important modifications from the loop-induced decay mode $h \to \gamma\gamma$, where nonstandard singly and doubly charged scalars may contribute significantly. Accordingly, the reference points are chosen such that the diphoton signal strength remains within the $2\sigma$ limit of the current experimental bound, $1.13 \pm 0.09$~\cite{CMS:2022dwd}.

\item {\bf Experimental bounds on the scalar masses:}
The direct search for the non-standard scalars has a strong impact on our model parameter space.  The direct search on the 
singly charged scalar in LEP II places a limit on $m_{H^\pm}\geq 80 $ GeV 
\cite{ALEPH:2013htx}. LHC has also searched for the decay of $ t \to H^+ b$~\cite{ATLAS:2024hya} and $ H^+ \to t \bar{b}$~\cite{ATLAS:2024rcu} depending on the mass $H^\pm$. However, for the type-II seesaw model, this coupling is $v_t/v_d$ suppressed, which does not affect our parameter space.

Collider searches for the non-standard neutral Higgs have also been done in detail at the LHC~\cite{ATLAS:2017uhp,ATLAS:2020tlo}. However, for this model, production of the nonstandard scalars is suppressed by $(v_t/v_d)^2$ and hence does not affect our parameter region. 

The search for doubly charged Higgs in the multi-lepton final state is done in ATLAS~\cite{ATLAS:2022pbd} assuming that the doubly charged Higgs decays promptly and masses $\in$[300:1080] GeV have been excluded. There is also a bound on doubly-charged Higgs from the production of same-sign W-boson pairs $\in$[200:220] GeV~\cite{ATLAS:2018ceg}. CMS has also searched for the doubly-charged scalar at 13 TeV~\cite{CMS:2017pet,CMS:2017fhs,CMS:2021wlt}. However, all searches assume that the scalar decays promptly. \\

$\bullet$~\underline{\textbf{Long-lived $H^{\pm\pm}$}:} 
The scenario changes significantly if the doubly charged scalar is \textbf{long-lived}. 
Depending on the triplet vev and the mass of the doubly charged Higgs, it can decay into a pair of same-sign dileptons, a pair of same-sign gauge bosons, or undergo a three-body decay via $H^{\pm\pm} \rightarrow W^{\pm}\left(W^{\pm}\right)^{*} \rightarrow W^{\pm} f \bar{f}^{\,\prime}$,
as discussed in Ref.~\cite{Kang:2014lwn}. 
By appropriately tuning the triplet vacuum expectation value 
$v_t$ and $m_{H^{\pm\pm}}$, the total decay width $\Gamma_{H^{\pm\pm}}$ of the doubly charged Higgs can be made very small 
$\Gamma_{H^{\pm\pm}} \lesssim 10^{-14}~\mathrm{GeV}$.
Such an extremely narrow width corresponds to a macroscopic proper decay 
length $c\tau \gtrsim \mathcal{O}(0.1~\mathrm{mm})$, rendering a long-lived particle(LLP) in 
collider detectors. A comprehensive analysis of this scenario has been performed in~\cite{Antusch:2018svb}. After imposing all the present constraints, a region of model parameter space remains viable with 
\begin{eqnarray} 
m_{H^{\pm\pm}} \in [89,\,200]~\mathrm{GeV}, ~~~~v_t \in [10^{-4},\,10^{-3}]~\mathrm{GeV},
\end{eqnarray}
where the doubly charged Higgs can evade existing bounds while showing a 
displaced or long-lived decay signature.
\end{itemize}

Taking into account all the constraints discussed so far, we choose our benchmark points as mentioned in Table~\ref{tab:bp}.  To satisfy all the constraints, we need the non-standard scalar masses to be degenerate. For our choice of $v_t$, the doublet-triplet CP even 
scalar mixing angle $\alpha $ is $ < 10^{-5}$. 
The values of the parameters for each benchmark with the corresponding proper decay lengths $c\tau$ of the doubly charged scalar and its branching ratios into muons are shown in Table~\ref{tab:bp}. The detailed expressions of the partial decay widths of the doubly charged scalar and the variation of its decay length with triplet vev is shown in Appendix \ref{app:partial decay length}.
As the mass of the doubly-charged scalar increases, the decay width increases, leading to a drop in $c\tau$. Once the $W^\pm W^\pm$ decay channel becomes kinematically accessible (third benchmark), the decay width increases drastically, resulting in a highly suppressed $c\tau$ and a relatively low branching ratio of $H^{\pm\pm}$ in the dimuon channel. These branching ratios play a crucial role in collider analysis as described in the following section.

\begin{table}[h!]
    \centering
    \begin{tabular}{|c|c|c|c|c|} \hline
         & $H^{\pm\pm}$ (GeV) & $v_t$ (GeV) & $c\tau$ (mm) & Branching Ratio to muons   \\ \hline
       BP1 & 100 &   $9.0\times10^{-4}$ & 26.2 & 19.5\%\\ \hline
       BP2 &   120 & $5.0\times10^{-4}$ & 8.6 & 17.2\%\\ \hline
       BP3 &   180 & $1.6\times10^{-4}$ & 0.14 & 6.1\%\\ \hline
          \end{tabular} 
    \caption{ Benchmark points satisfied by all the theoretical and experimental constraints.}
    \label{tab:bp}
\end{table}

\section{Collider Analysis }~\label{result}
Our main goal is to explore the long-lived nature of the doubly charged scalars through their leptonic decay at 
future leptonic colliders. For this purpose, we focus on the model parameter space that provides $M_{H^{\pm\pm}}\sim 100-200$ GeV 
with some fine-tuned choices of $v_t$. The choice of model parameters is such that $H^{\pm\pm }$ predominantly decays to 
$\mu^\pm \mu^\pm$ final state. At the lepton colliders, such a doubly charged scalar can be produced 
either minimally, in a pair (process 1) via the canonical Drell-Yan process or in a pair with missing energy (process 2), which includes the vector boson fusion (VBF) contribution.
 The mass of the final particles being in the range of $[100-200]$ GeV, the Drell-Yan cross-section is considerable to probe at a lepton collider with $\sqrt{s}\sim[200-400]$ GeV, while the cross-section for process 2 is an order of magnitude smaller compared to it. Therefore, ILC, planned to operate in the ballpark of this energy, can offer a good environment for probing the Drell-Yan production. But with increasing CM energy, 
$\sqrt{s}\sim\mathcal{O}(1)$TeV, the Drell-Yan becomes smaller due to s-channel suppression, while the process 2 can dominate over the Drell-Yan production cross-section. The proposed muon collider with a 10 TeV CM energy appears to be an ideal choice for searching for the latter channel. In the latter case, doubly charged scalars 
are associated with either two forward-moving $\mu^+ \mu^-$ or $\nu_\mu {\bar \nu_\mu }$, depending on the characteristic of the vector boson involved in the fusion process. However, due to the complex structure of the underlying framework 
and the large particle spectrum, one has to consider other sub-processes involving additional scalars of the model in association with the vanilla VBF contributions for the second production mode. In both cases, our final targets will be four charged muons 
emanating from the decays of two doubly charged scalars. In our analysis, we focus on the Drell-Yan process in the ILC environment
and the associated production of $H^{\pm\pm} H^{\mp\mp}$ with forward leptons ($\mu^\pm, {\rm or}~ \nu_\mu {\bar \nu_\mu})$ at a muon collider.

The Drell-Yan signal cross-section at ILC  and the cross-section for process 2 at the muon collider, considering the muonic decay of the doubly charged Higgs, for three benchmark points are mentioned in Table\,\,\ref{tab:cross_ilc} for different CM energies, $\sqrt{s} =500$ GeV for the first process (DY) and $\sqrt{s}=10$ TeV for the second process. As discussed earlier, for low energies, Drell-Yan performs better, whereas for higher energies, process 2 is a better bet.
The type II seesaw model specific details enter the analysis via the branching ratio of the doubly charged Higgs to the leptons, which is controlled by the triplet vev $v_t$. The branching ratios of the doubly charged Higgs to muons for the benchmark $v_t$s are mentioned in Table\,\ref{tab:bp}. 

To provide an estimate of the cross-sections at other proposed colliders, in Table.\ref{tab:cross_othercollider}, we have reported the signal and background cross sections for process 1 at $\sqrt{s} = 365$ GeV ($t{\bar t}$ threshold) which is one of the propsed CM energy of FCC-ee~\cite{FCC:2018byv} and $\sqrt{s}=3$ TeV which is the proposed CM energy for CLIC proposal \cite{Brunner:2022usy}, and process 2 at $\sqrt{s} = 3$ TeV corresponding to a proposed CM energy quoted in the earlier muon collider TDR~\cite{MuonCollider:2022xlm}.

\begin{table}[h!]
    \centering
    \begin{tabular}{|c|c|c|c|c|} \hline
    
    Cross-section & BP1 & BP2 & BP3 & SM Background \\ \hline
     $\sigma^{\rm process~1}_{\rm ILC}$  (fb) & 12.18 & 7.18 & 0.62 & 0.64 \\ \hline
   $\sigma^{\rm process~2}_{\rm IMCC}$  (fb) & 0.09 & 0.06 & 0.006 & - \\ \hline 
    \end{tabular}
    \caption{Production cross-section of the signal processes 1 and 2 for our chosen benchmark points and the corresponding SM background at the ILC at $\sqrt{s}=$ 500 GeV and muon collider at $\sqrt{s}=$ 10 TeV.}
    \label{tab:cross_ilc}
\end{table}

\begin{table}[h!]
    \centering
    \begin{tabular}{|c|c|c|c|c|c|} \hline
    
   $\sqrt{s}$& Cross-section & BP1 & BP2 & BP3 & SM Background \\ \hline
    365 GeV& $\sigma^{\rm process~1}_{e^+ e^-}$  (fb) & 17.40 & 9.99 & 0.01 & 1.06 \\ \hline
    3 TeV& $\sigma^{\rm process~1}_{e^+ e^-}$  (fb) & 0.005 & 0.009 & 0.006 & 0.01 \\ \hline
   3 TeV &$\sigma^{\rm process~2}_{\mu^+ \mu^-}$  (fb) & 0.09 & 0.06 & 0.005 & - \\ \hline 
    \end{tabular}
    \caption{Production cross-section of the signal processes 1 and 2 for our chosen benchmark points and the corresponding SM background at the electron collider for $\sqrt{s} = 365\,{\rm GeV}$ (FCC-ee proposal), $\sqrt{s}= 3\,{\rm TeV}$ (CLIC proposal) and $\mu^+\mu^-$ collider for $\sqrt{s}=3\,{\rm TeV}$.}
    \label{tab:cross_othercollider}
\end{table}

\subsection{Pair production of the doubly-charged Higgs via Drell-Yan process at the ILC}

In this section, we discuss the prospects for detecting doubly charged LLPs produced in the Drell-Yan process at the 500 GeV ILC, followed by their decay to same-sign muon pairs:
\begin{eqnarray}
e^+ e^- \to H^{++} H^{--}, H^{\pm\pm} \to \mu^{\pm}\mu^{\pm}.
\label{eq:ilc}
\end{eqnarray}
The SM process that can mimic the signal comes from the di-boson background. To take this contribution into account, we have considered inclusive generation of the background process  $e^+ e^- \to \mu^+ \mu^+ \mu^- \mu^-$.
While generating both the signal and the SM background events in $\tt{MadGraph5@NLO}$~\cite{Alwall:2014hca}, we employ the following pre-selection cuts used for ILC :
\begin{equation}
\tt{\left|\eta(\ell)\right| < 2.5;~~~ p_{T}(\ell) > 10~GeV;~~~ \Delta R_{\ell\ell} \geq 0.4}.
\label{eq:preselection}
\end{equation}
To justify the $\Delta R$ cut used in the pre-selection, the corresponding distribution has been shown in the Appendix.\ref{app:delta R distribution}. 
The events are then passed to $\tt{PYTHIA8}$\cite{Sjostrand:2014zea} for hadronization, and finally, detector simulation is performed using $\tt{DELPHESv3}$\cite{deFavereau:2013fsa} with the help of the ILD card~\cite{Behnke:2013lya}. 

\begin{figure}
    \centering
     \includegraphics[scale=0.38]{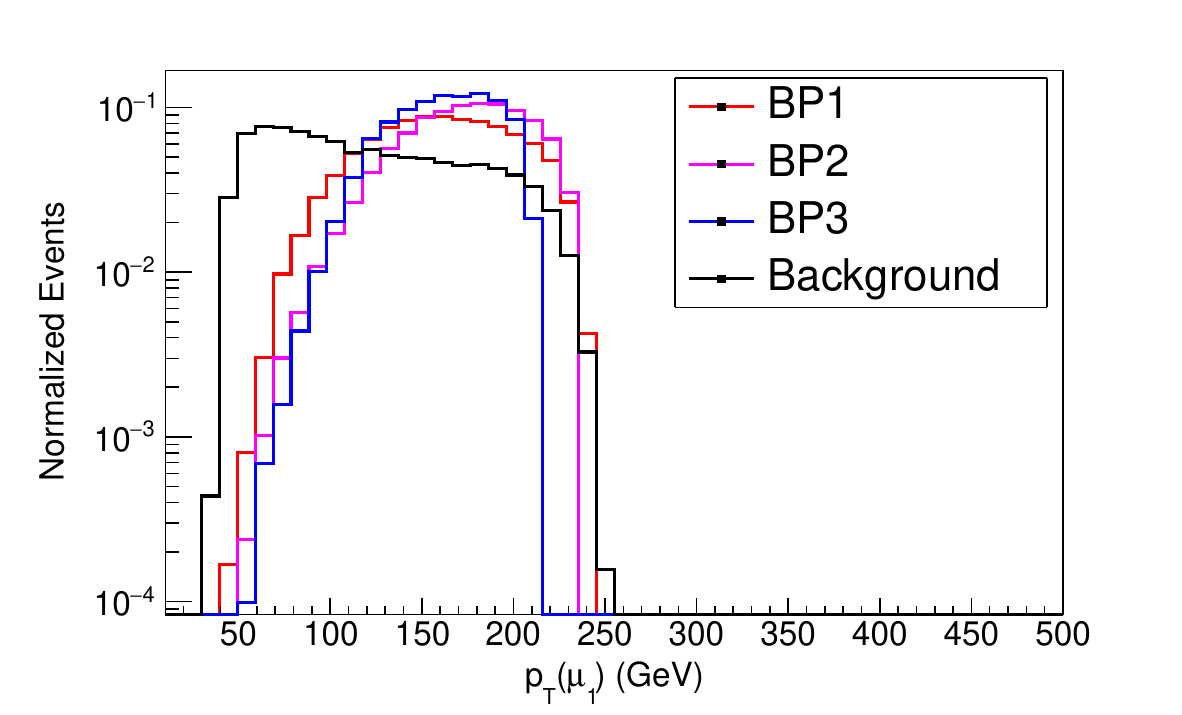}
      \includegraphics[scale=0.38]{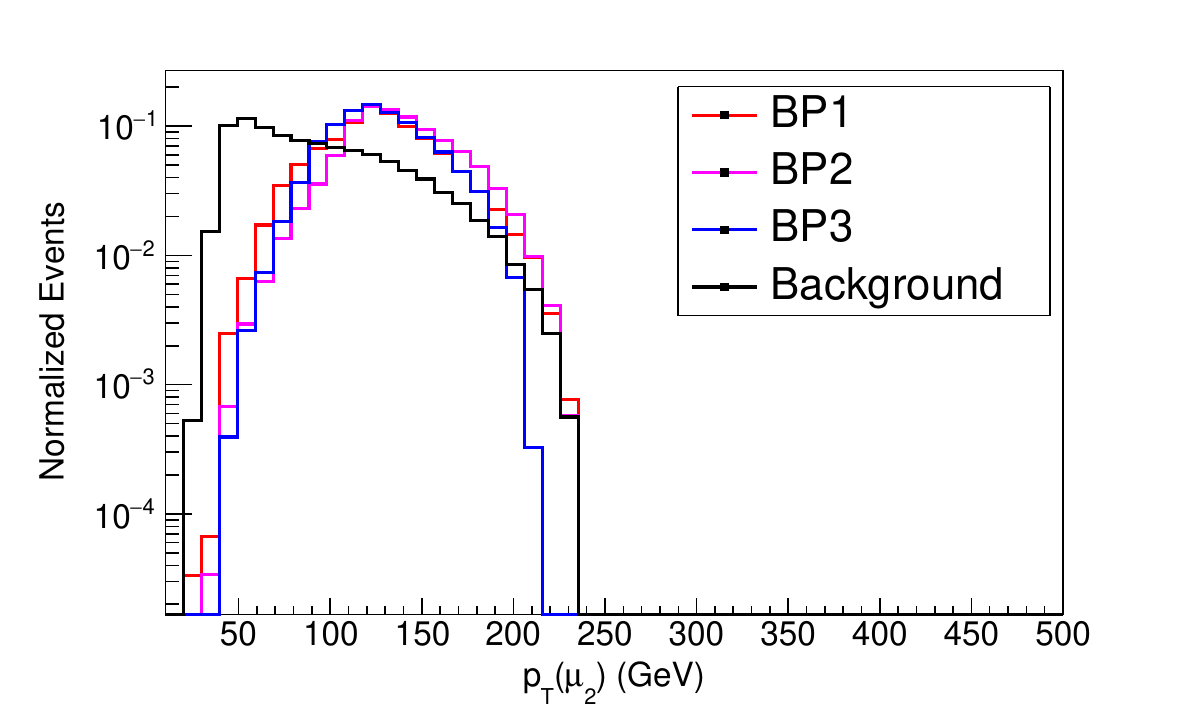}\\
       \includegraphics[scale=0.38]{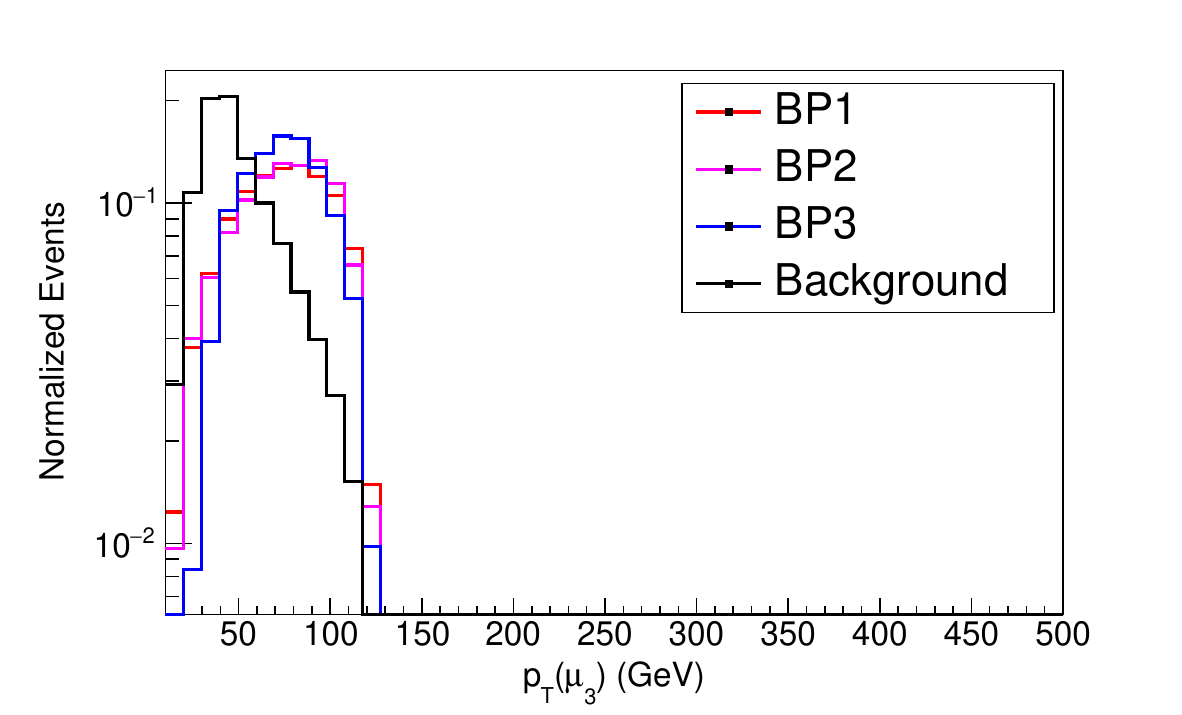}
        \includegraphics[scale=0.38]{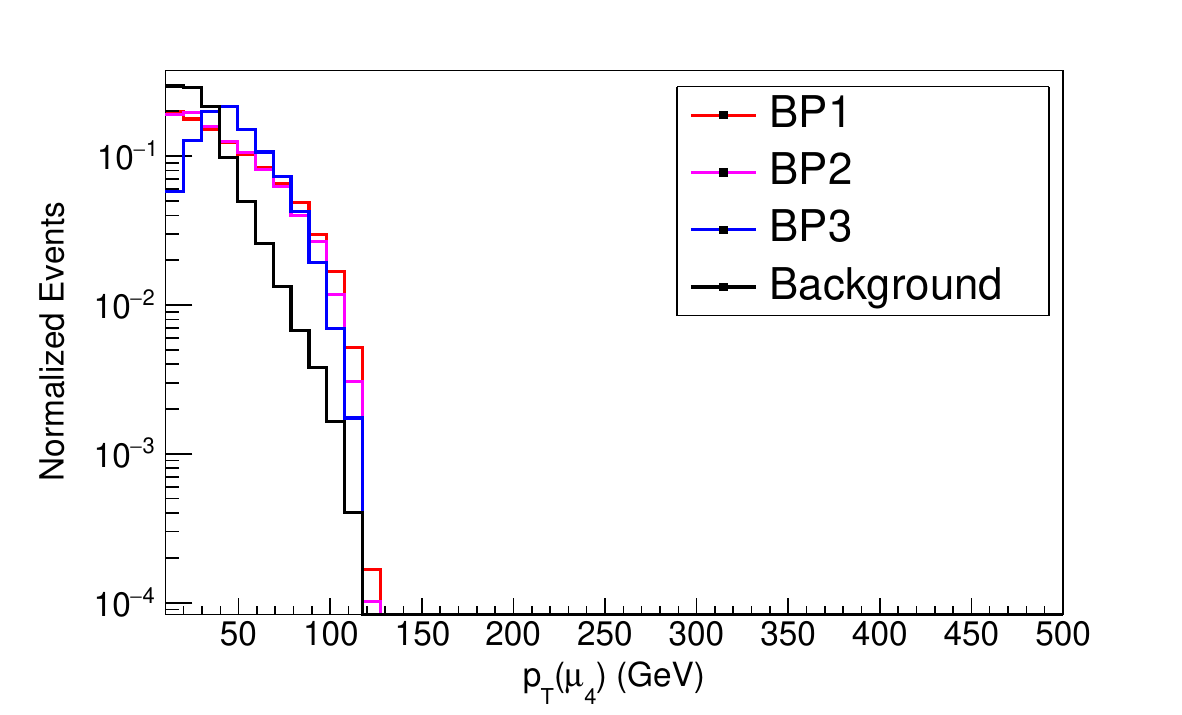}
\caption{Normalized transverse momentum distributions of the leading and sub-leading muons in signal events are shown for benchmark points BP1, BP2, and BP3. For comparison, the corresponding Standard Model background distributions are also included. Results are presented for the ILC at $\sqrt{s} = 500$ GeV with integrated luminosity, $\mathcal{L}_{\rm int}=4 ab^{-1}$. 
}
    \label{fig:ilc_ptmu}
\end{figure}

\begin{figure}
    \centering        
         \includegraphics[scale=0.38]{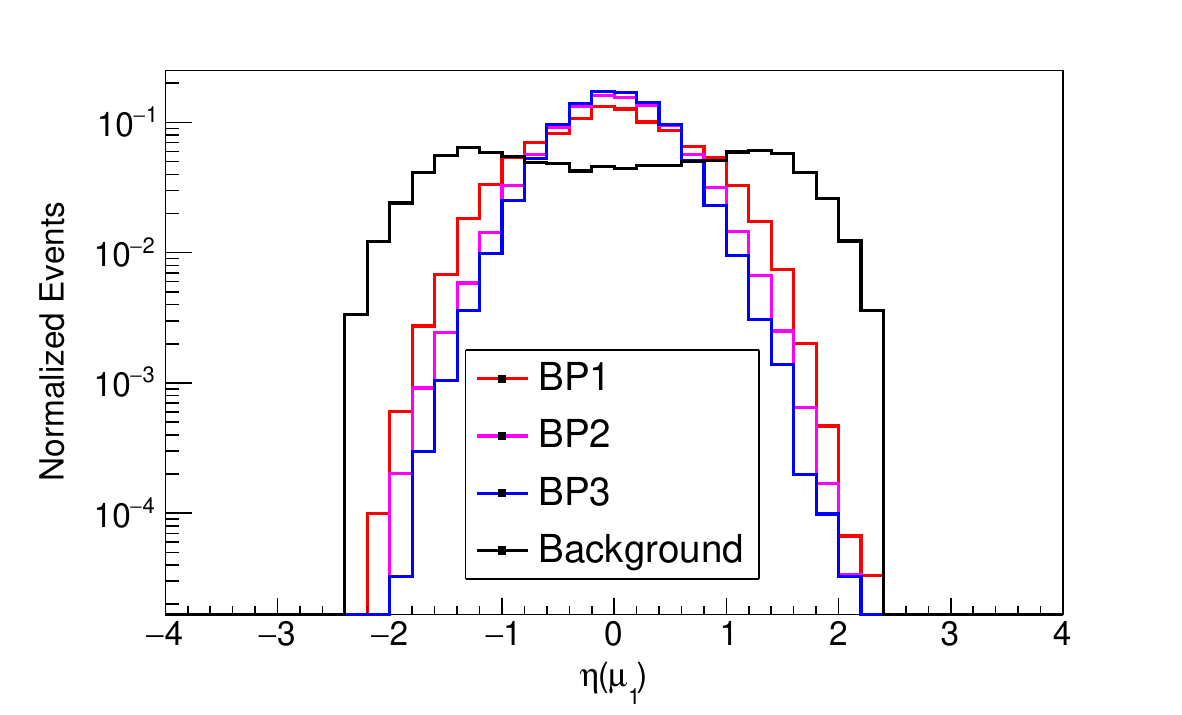}
         \includegraphics[scale=0.38]{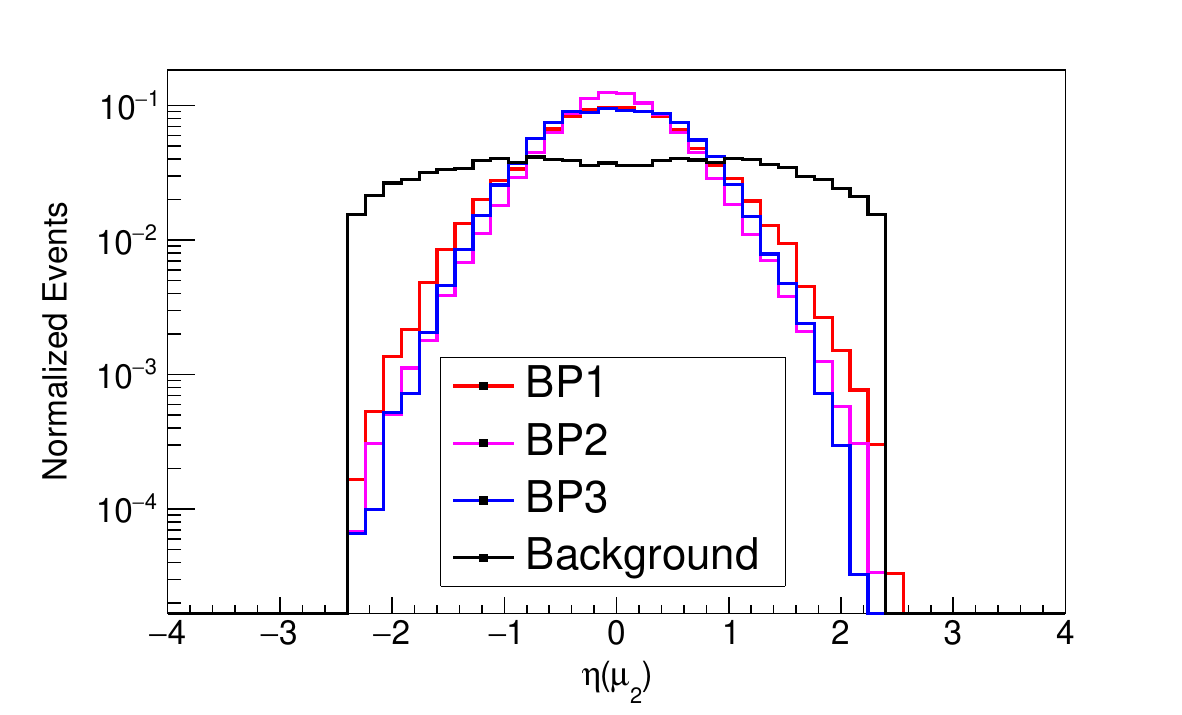}\\
         \includegraphics[scale=0.38]{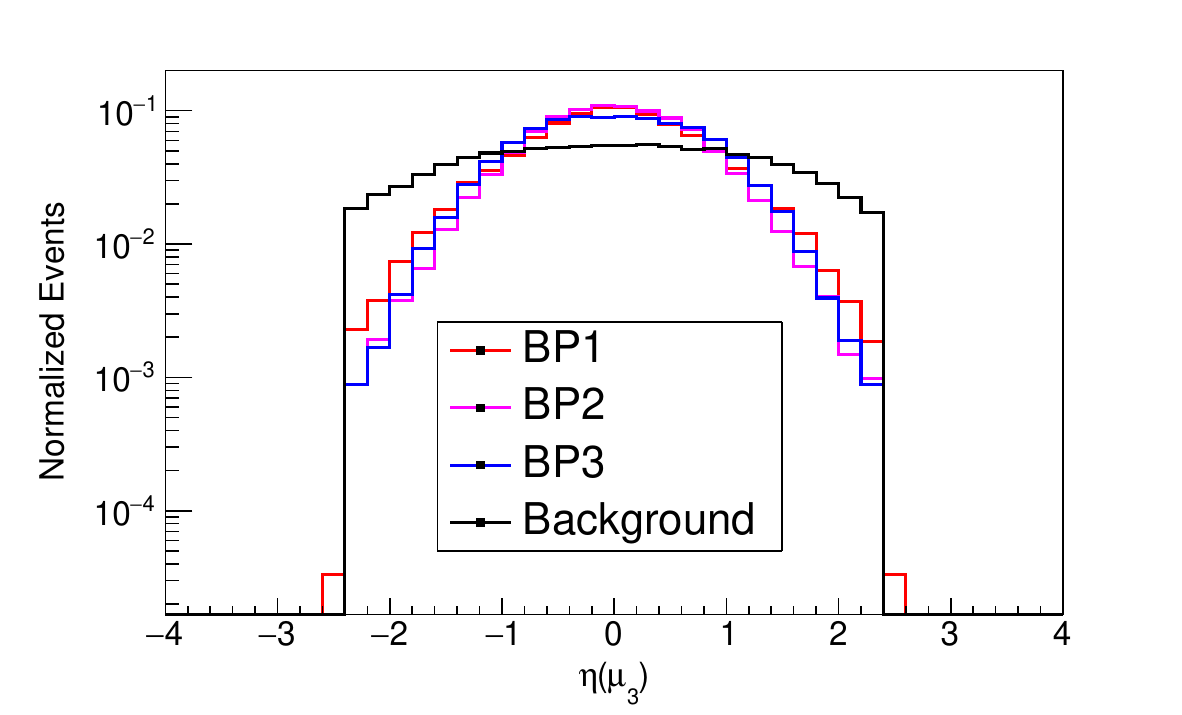}
         \includegraphics[scale=0.38]{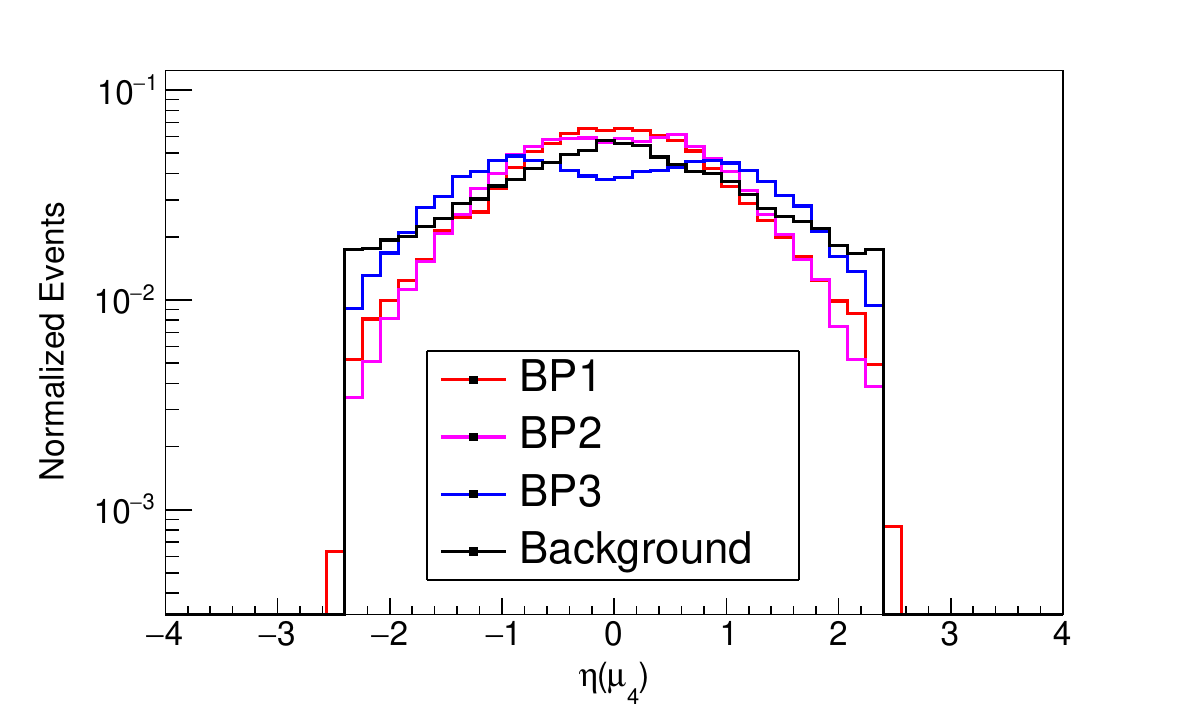}
\caption{ Normalized pseudo-rapidity distributions of the leading and sub-leading muons in signal events are shown for benchmark points BP1, BP2, and BP3. For comparison, the corresponding Standard Model background distributions are also included. Results are presented for the ILC with the same CM energy and luminosity as Figure 1.}

    \label{fig:ilc_etamu}
\end{figure}

\begin{figure}[h]
    \centering
    \includegraphics[scale=0.38]{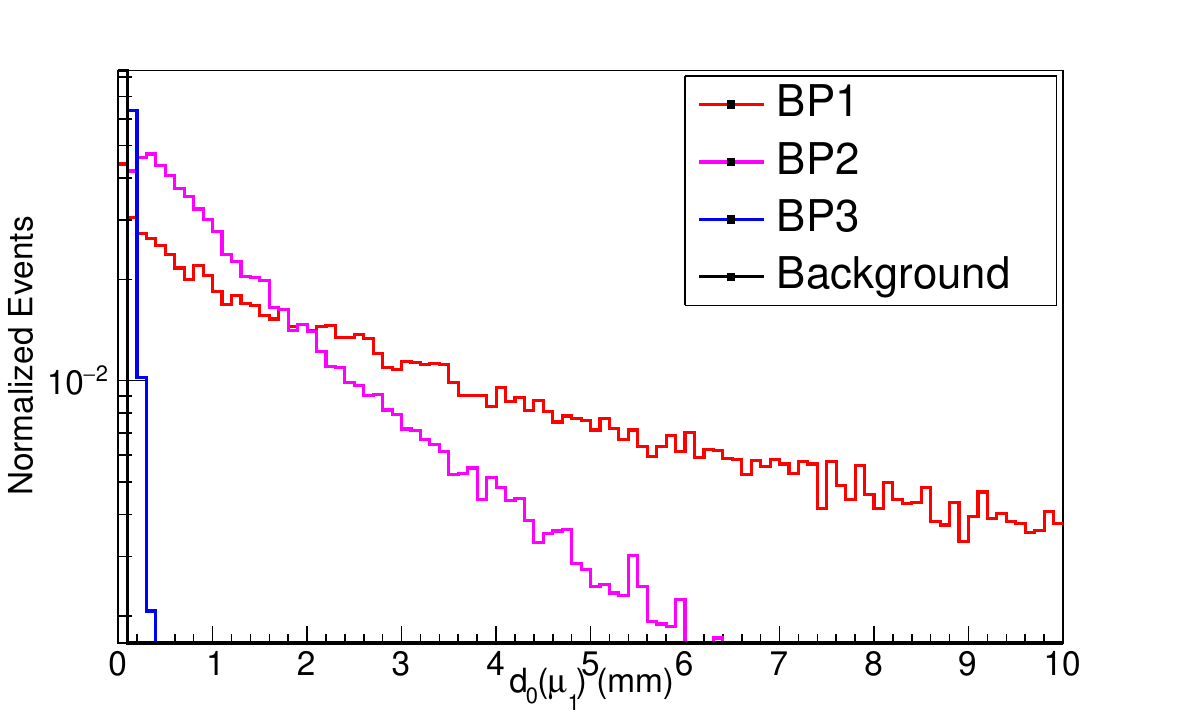} 
    \includegraphics[scale=0.38]{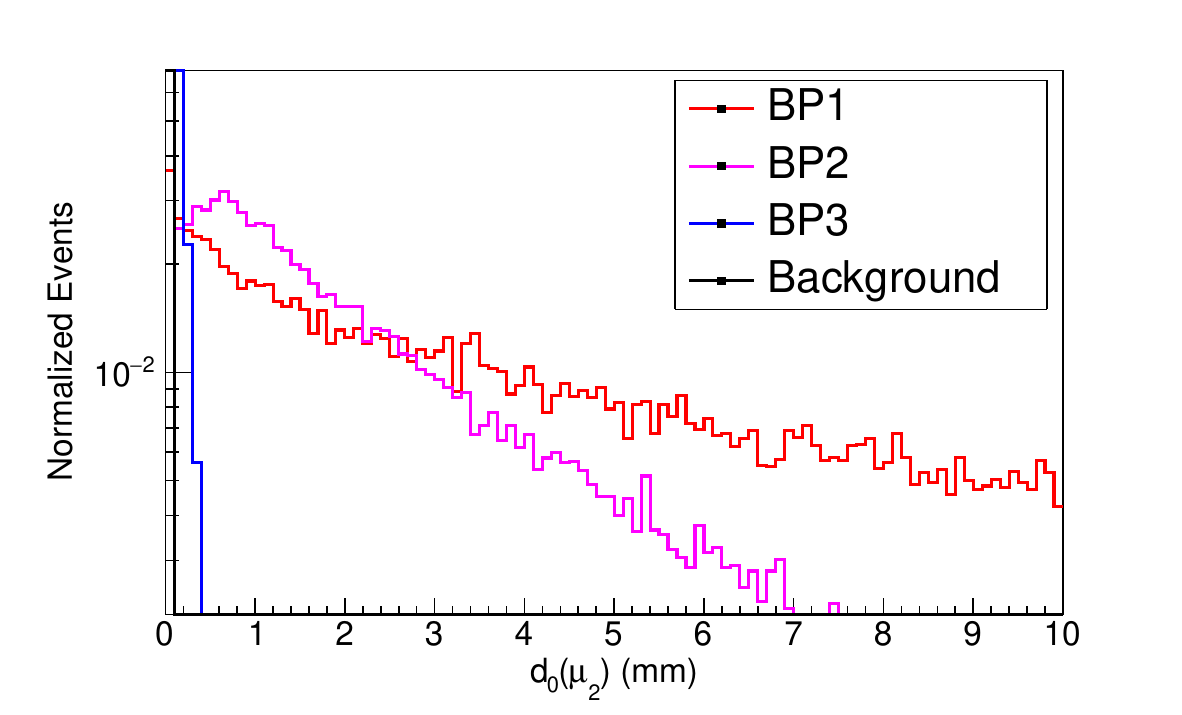} \\
    \includegraphics[scale=0.38]{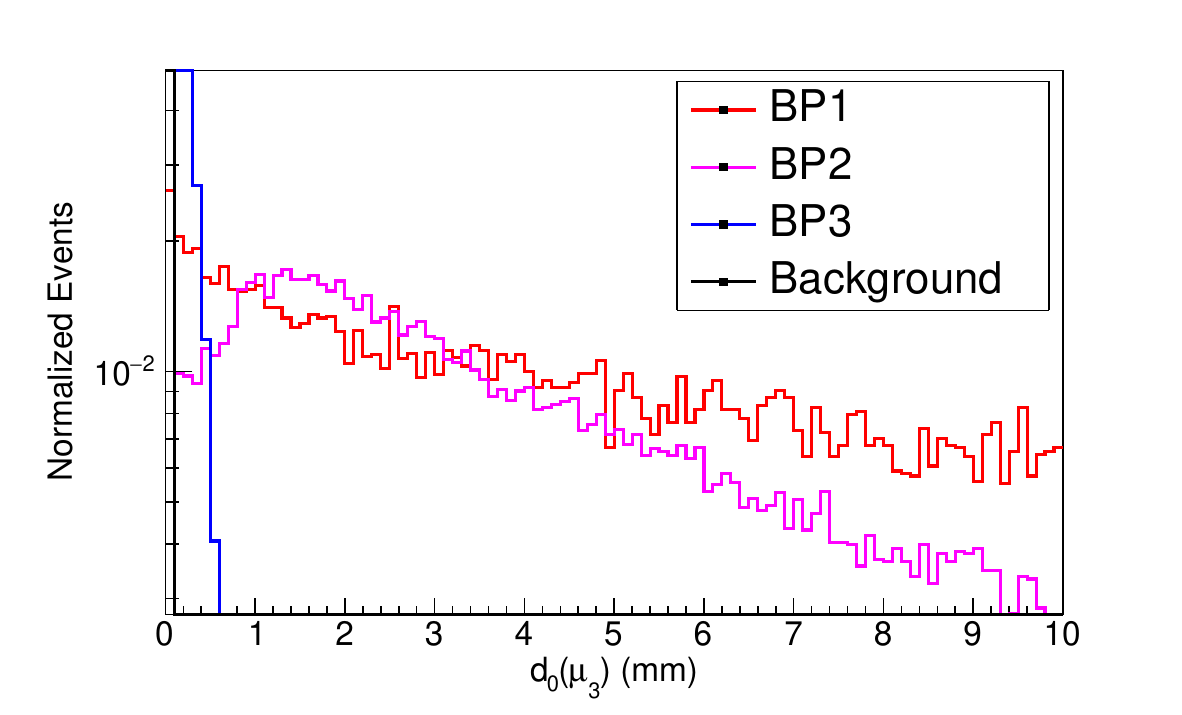} 
    \includegraphics[scale=0.38]{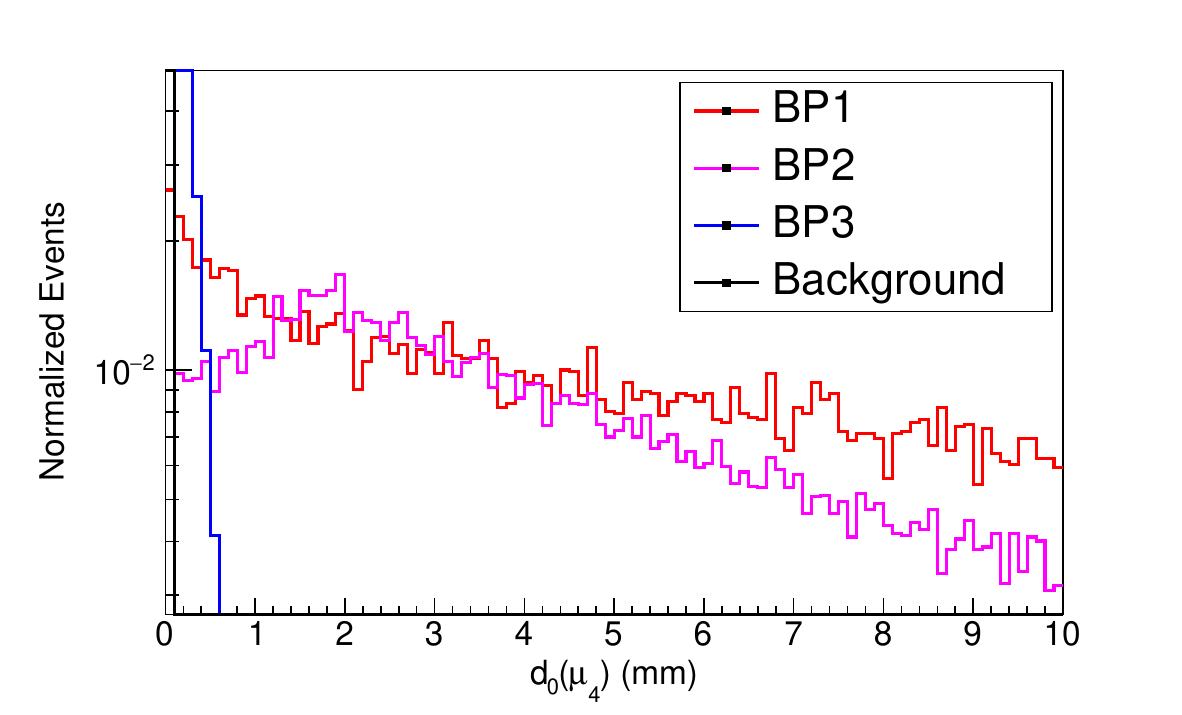} 
    
\caption {Normalized impact parameter $\mid d_0 \mid $ distributions of the leading and sub-leading muons in signal events as well as the SM background. The CM energy and luminosity considered are taken to be the same as in Figure 1.   }
    \label{fig:ilc_track}
\end{figure}

\noindent In addition to the basic cuts described in Eq.~\ref{eq:preselection}, we employ the following set of cuts to enhance the signal over the background:
\begin{itemize}
    \item C1-1: We demand that the signal has exactly 4 muons and reject any other leptons, photons, and jets.
    \item C1-2: In Fig.~\ref{fig:ilc_ptmu}, we plot the normalized distribution of the transverse momentum of the leading and sub-leading muons. For the signal, muons come from $H^{\pm\pm}$ and thus tend to peak in a slightly higher $p_T$ bin, since the doubly-charged scalar mass is not that heavy. In the background, the leptons come from the gauge bosons. We put a moderate cut on the muon $p_T (\ell) > 30$ GeV.
    \item C1-3: We present the pseudorapidity distribution for both the signal and the background in Fig.~\ref{fig:ilc_etamu}. Since for the signal, the leptons are produced centrally, we put a cut of $|\eta_{\ell}| < 2.0$ to reduce the background.
    \item C1-4: In the context of LHC to search for a displaced vertex, an important variable, the impact parameter $d_0$, is used in the literature to remove SM backgrounds ~\cite{ATLAS:2012qqw,Cerdeno:2013oya,Abdallah:2018gjj}. Now, a track of a doubly charged scalar can be a very interesting signature for this kind of signal~\cite{Lu:2024ade}. However, the reconstruction of that track in such an environment and the corresponding analysis are beyond the scope of our analysis and can be explored in the near future. The
resolution of the vertices in the pixel tracker for both ATLAS and CMS detectors is of the
order of hundreds $\mu$m~\cite{Kobayashi:2019gdv,CMSTrackerGroup:2025eoc} and usually a cut of a few mm is imposed to suppress the SM background~\cite{Antusch:2018svb}. Compared to LHC, the sensitivity of the ILC pixel is of the order of a few $\mu$m~\cite{ILCInternationalDevelopmentTeam:2022izu}, and one can expect that a little bit of displacement from the primary interaction point will be detected more efficiently at the ILC. In Fig.~\ref{fig:ilc_track}, we show the normalized distribution of the impact parameter $d_0$ of the leading and sub-leading muons. Since the muons are appearing from the long-lived $H^{\pm\pm}$, they are displaced from the primary vertex, and as a result, the daughter muons are also displaced. The transverse impact parameter $d_0$ exhibits an approximate dependence $\frac{1}{p_T}$ arising from the curvature of the track in the magnetic field~\cite{Han:2005mu}; consequently, benchmarks with higher muons $p_T$ lead to smaller transverse impact parameters, as illustrated in Fig.~\ref{fig:ilc_track}.
 In the background, muons come from the $W^{\pm}, Z$ prompt boson and are produced at the primary vertex.  Finally, we demand 4 displaced muon tracks with  $|d_0| > 0.01, 0.1 (1)$ mm to remove the SM background completely. Here we would like to emphasize that even the signal muon with the smallest $d_0$ is demanded to be greater than $>$ 0.1 mm.  To independently validate the displaced signal topology, the $H^{\pm\pm} \to \mu^\pm\mu^\pm$ decay vertices are reconstructed from the four final-state muon tracks using their points of closest approach. A common origin is ensured by requiring a distance of closest approach (DCA)
 $< 1\text{ mm}$ between the same-sign paired tracks. Finally, the reconstructed vertices must lie within the detector's fiducial volume and satisfy a minimum transverse decay length $\rho > X$, where $X = 0.01, 0.1$ and 1 mm.
\end{itemize}

Based on the cut-flow mentioned above, we quote the number of events at 4 $\rm ab^{-1}$ luminosity at the ILC in Table~\ref{tab:cutflow_ilc}. For the purpose of comparison between the efficiency of the $d_0$ and $\rho$ cut, we have shown the number of events after applying the $d_0$ cut and $\rho$ as final cuts in Table~\ref{tab:comp_ilc}. The reach of the five signal events using the $d_0$ cut corresponds to an integrated luminosity of
$\sim 3~\mathrm{fb}^{-1}$ for BP1 and BP2, increasing to
$\mathcal{O}(10^{3})~\mathrm{fb}^{-1}$ for BP3, as shown in Table~\ref{tab:5event_ilc}.

\begin{table}[h!]
    \centering
      \resizebox{\textwidth}{!}{ 
    \begin{tabular}{|c|c|c|c|c|c|c|} \hline
     Benchmark points  & C1-1 & C1-2 & C1-3 & \multicolumn{3}{c|}{C1-4} \\ \hline
     & & & & $d_0 >$0.01 mm & $d_0 >$0.1 mm & $d_0 >$1 mm \\ \hline
      BP1 & 11573 & 5980  & 5980  &  5891  & 5504  & 3302  \\ \hline
       BP2  & 12504 & 7711  & 7639  & 5239  & 5034  & 2360  \\ \hline
       BP3 & 1338  & 1094  & 1077 & 553  & 17 & 0 \\ \hline
        SM Background & 1290  & 536  & 460  & 0 & 0 & 0 \\ \hline
    \end{tabular}}
    \caption{Cut flow table showing the number of signal events
    at the ILC. The considered CM energy and the integrated luminosity are taken to be the same as in Figure 1.  
    The last three sub-columns show the final signal events after three choices of $d_0$ cut.}

    \label{tab:cutflow_ilc}
    \end{table}

\begin{table}
    \centering
    \begin{tabular}{|c|c|c|c|c|c|c|} \hline
     Benchmark points  & \multicolumn{2}{c|}{$> 0.01$ mm } & \multicolumn{2}{c|}{$> 0.1$ mm}  & \multicolumn{2}{c|}{$> 1$ mm} \\ \hline
     & $\rho$ & $d_0 $ & $\rho $ & $d_0 $ & $\rho$ & $d_0 $  \\ \hline
     BP1& 5980 & 5891 & 5861 & 5504 & 5088 & 3302  \\ \hline
     BP2& 5262 & 5239 & 5251 & 5034 & 4588 & 2360  \\ \hline
    BP3& 826 & 553 & 113 & 17 & 0 & 0  \\ \hline
\end{tabular}
    \caption{Comparison of the number of signal events
    at the ILC after $d_0$ and $\rho$ cut.}  
    \label{tab:comp_ilc}
    \end{table}
    
    \begin{table}
    \centering
    \begin{tabular}
        {|c|c|} \hline
       Benchmark Points & Luminosity required for 5 events \\
       & after $|d_0| >0.1$mm cut (${\cal{L}}_{5}$) (fb$^{-1}$)\\ \hline
        BP1 &  3.6 \\ \hline
        BP2 & 4.0 \\ \hline
        BP3 & 1196.5 \\ \hline
    \end{tabular}
     \caption{Luminosity required to see 5 signal events at the ILC.}
    \label{tab:5event_ilc}
\end{table}

\begin{figure}
    \centering 
     \includegraphics[scale=0.4]{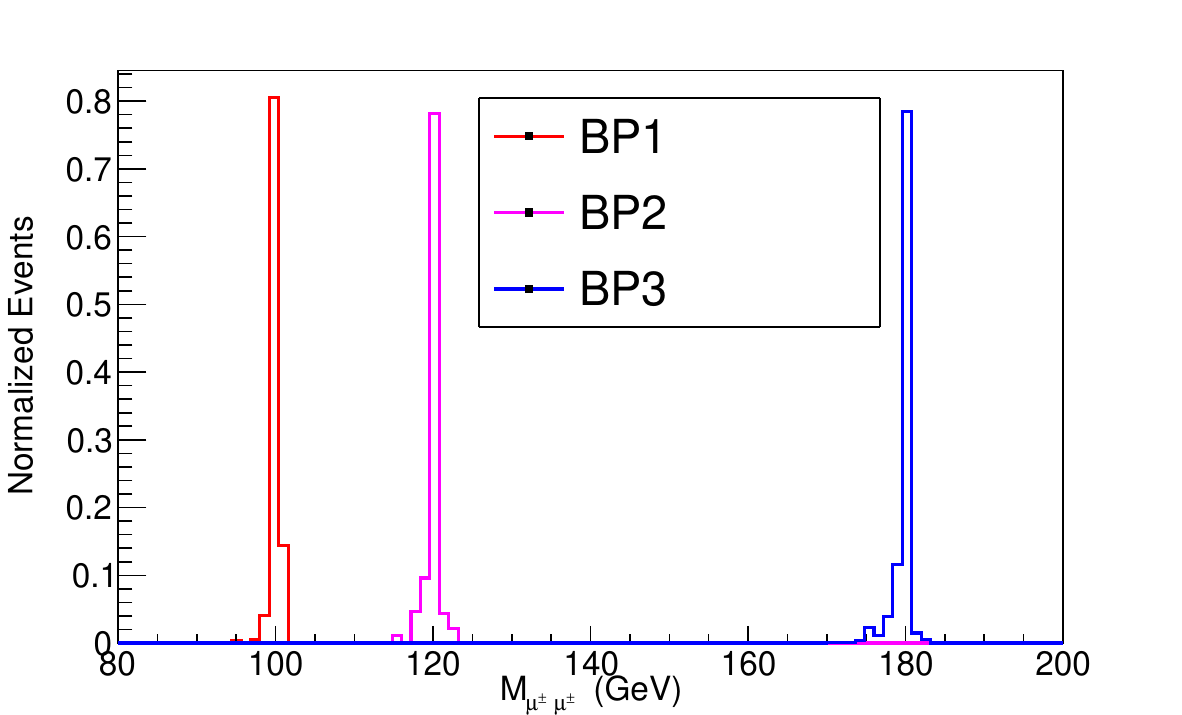}
  \caption{Normalized invariant mass distribution for same-sign dimuon pair for 3 chosen benchmark points at the ILC with the same CM energy and luminosity as in Figure 1. }
    \label{fig:ilc_inv}
\end{figure}

Now, ILC can also be used as a discovery machine for the doubly-charged Higgs. In Fig.~\ref{fig:ilc_inv}, we show the invariant mass distribution of the dimuons of the same-sign $M_{\mu^\pm \mu^\pm}$ after killing the SM background and see that for our chosen benchmark points, the invariant mass peaks at the exact bin with very little spread.

\begin{figure}
    \centering   
    \includegraphics[scale=0.35]{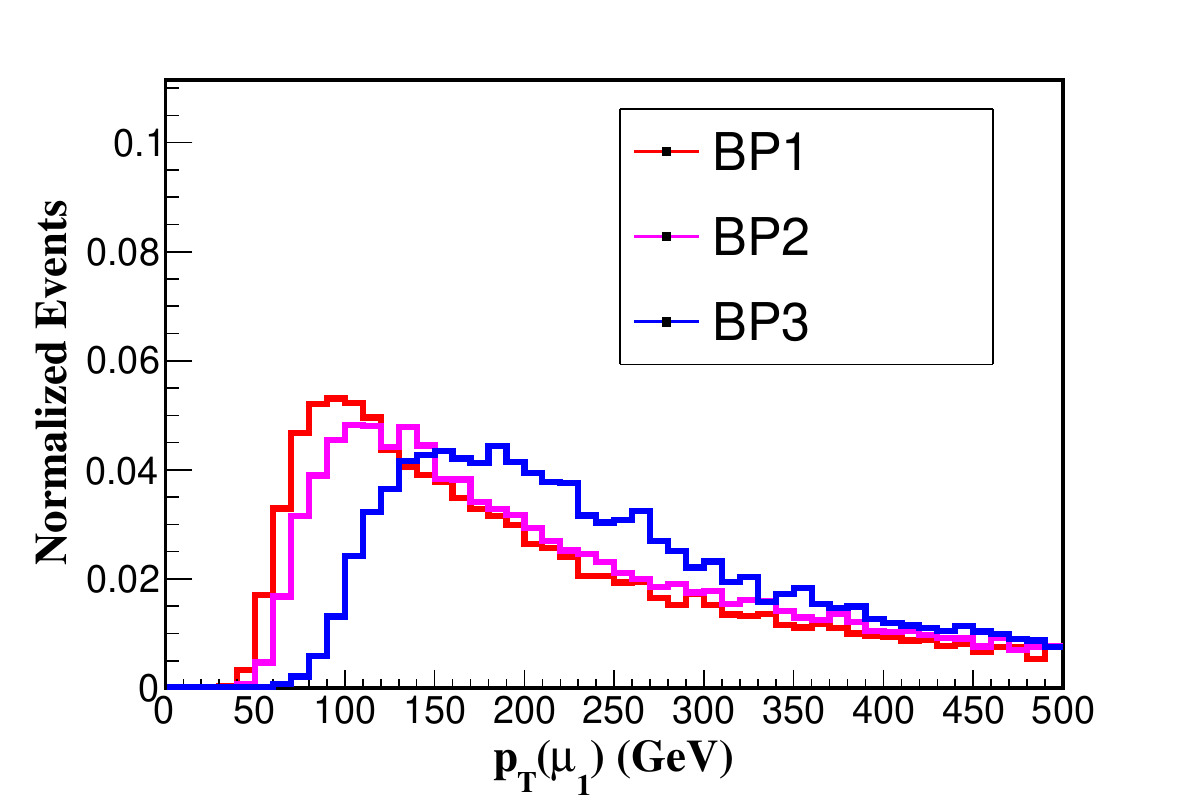}
    \includegraphics[scale=0.35]{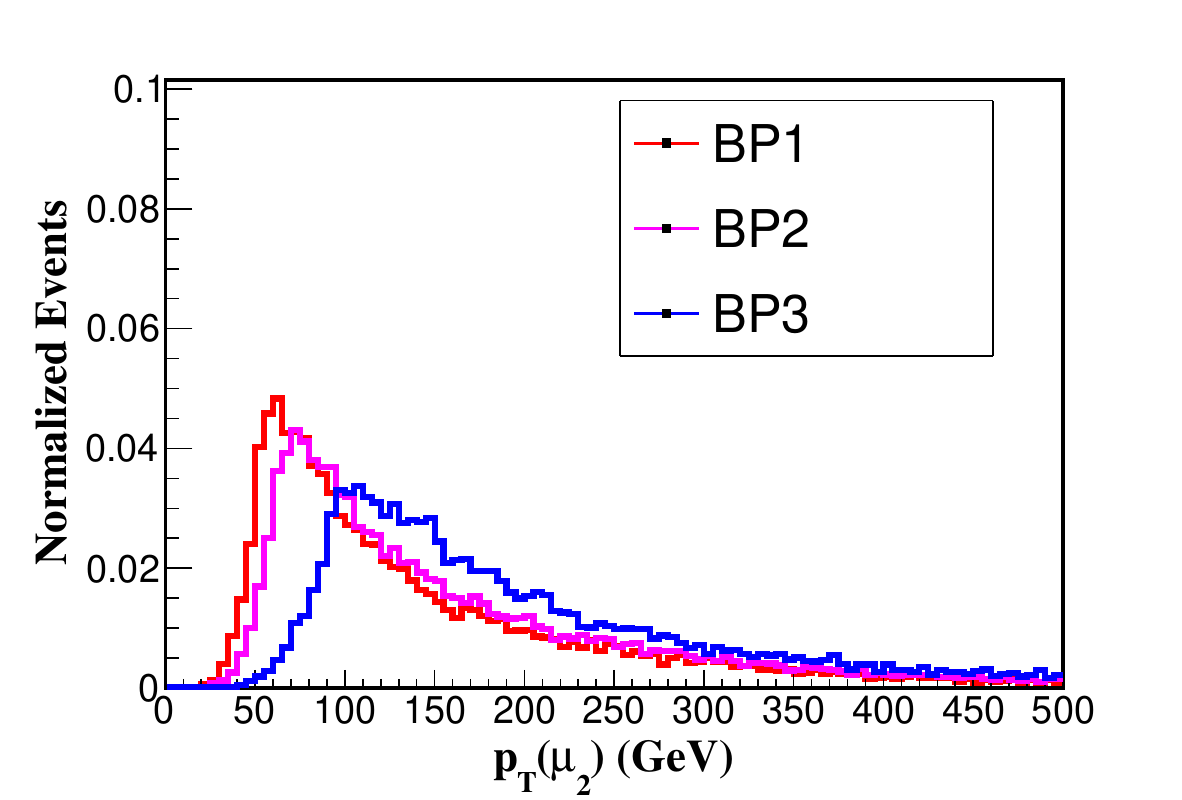}\\
    \includegraphics[scale=0.35]{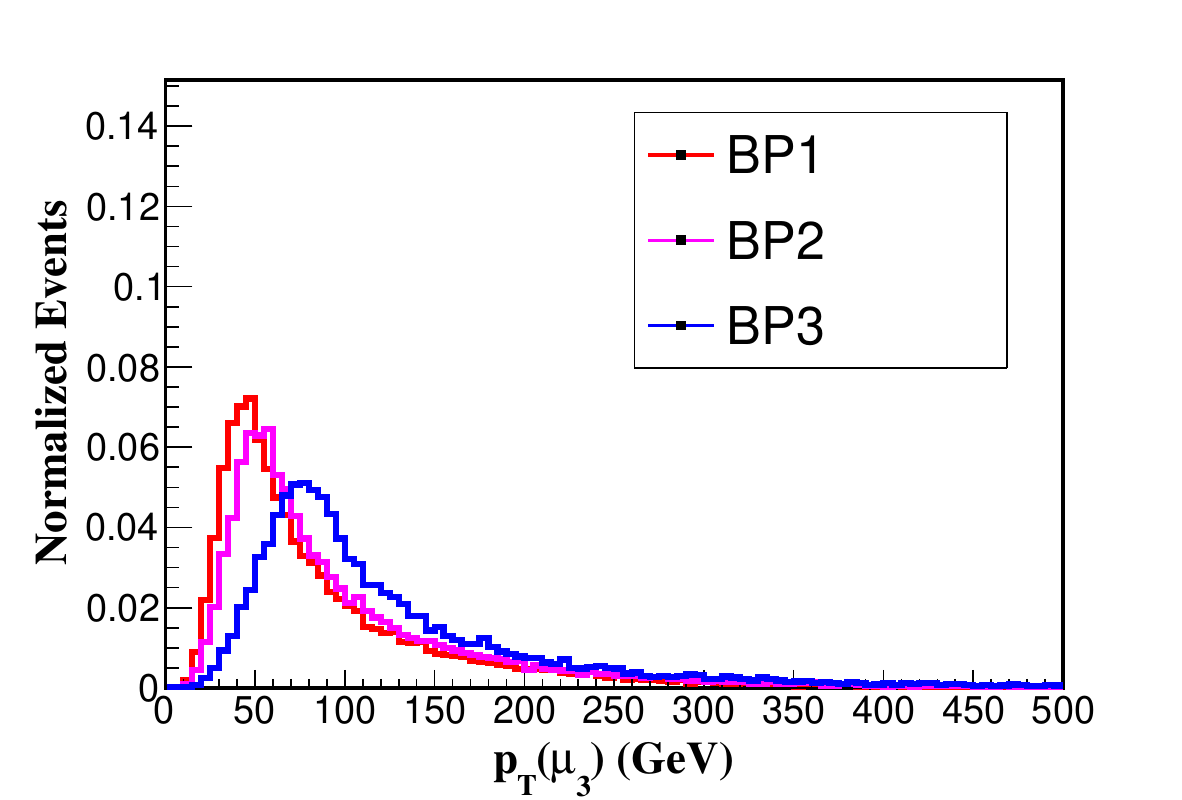}
    \includegraphics[scale=0.35]{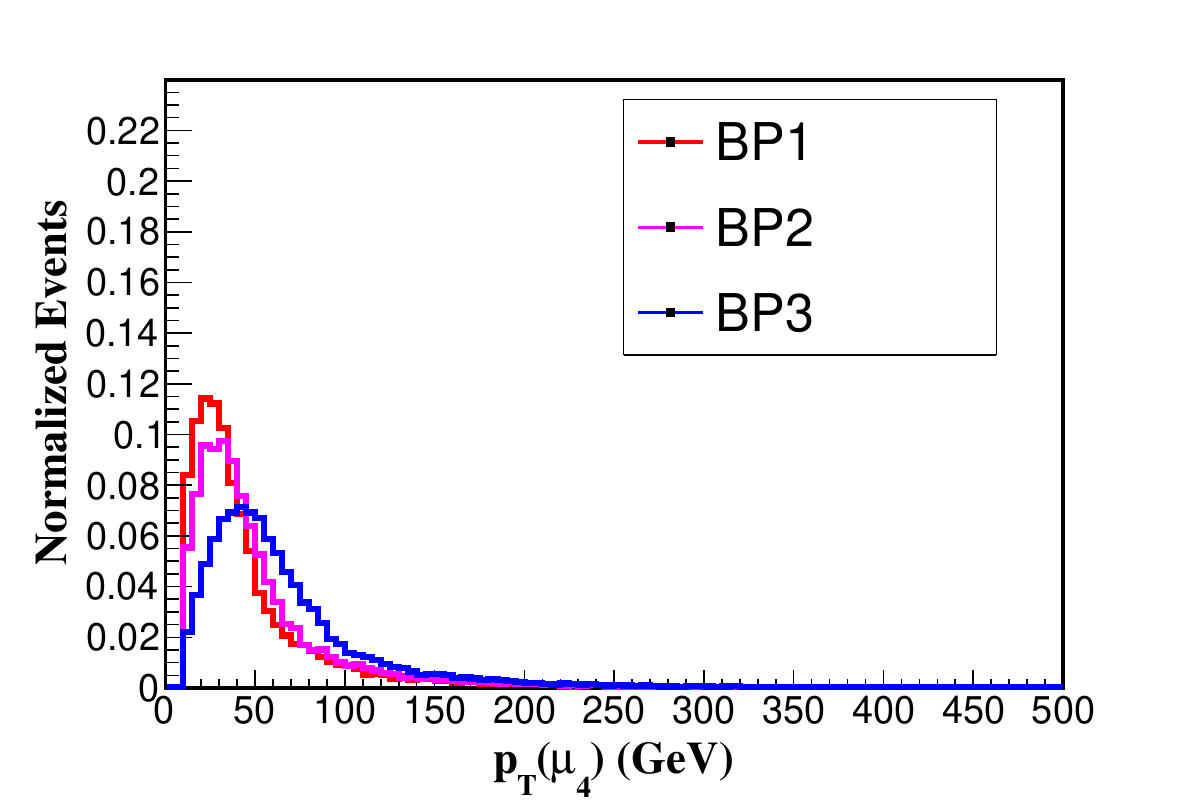}
\caption{Normalized transverse momentum distributions of the leading and sub-leading muons in signal events are shown for benchmark points BP1, BP2, and BP3. Results are presented for the muon collider at $\sqrt{s} = 10$ TeV with $\mathcal{L}_{\rm int}=10 ~{\rm ab}^{-1}$.}
    \label{fig:imcc_ptmu}
\end{figure}

\begin{figure}
    \centering
    $$
\includegraphics[width=0.5\linewidth]{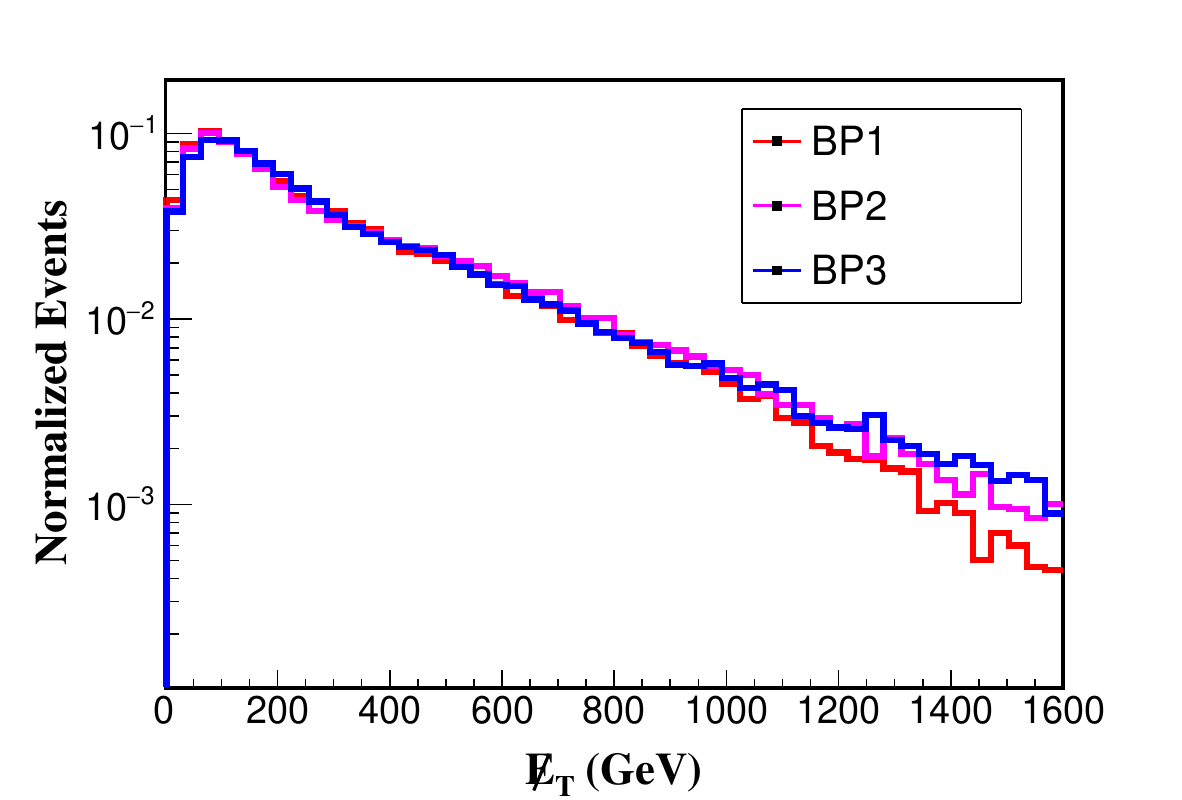}
\includegraphics[width=0.5\linewidth]{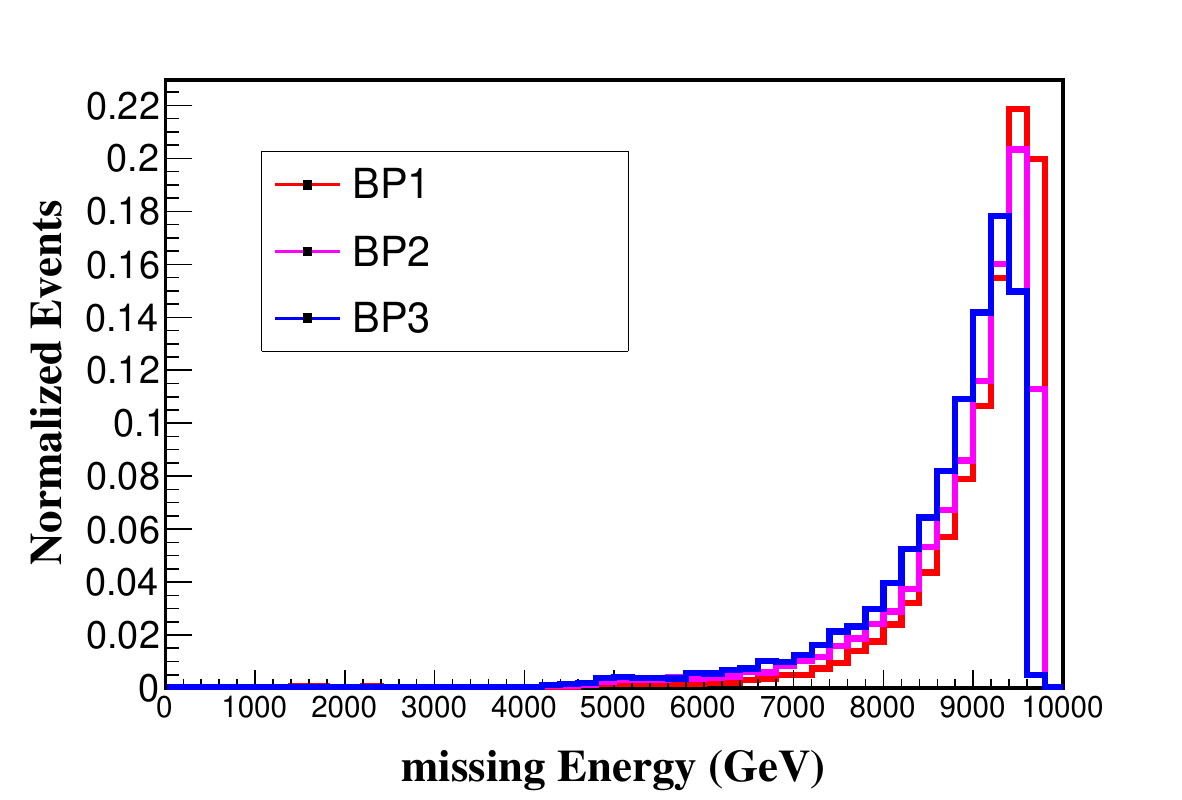}
$$
    \caption{Normalized transverse missing energy distributions in signal events are shown for benchmark points BP1, BP2, and BP3 for the muon collider with the same CM energy and luminosity as in Figure 5.}
    \label{fig:imcc_met}
\end{figure}

\subsection{Pair production of doubly charged Higgs at muon collider with missing energy} 
For the pair production of doubly charged Higgs, when the minimal signal is just the Higgs pair, the next-to-minimal signal appears to be a pair of charged Higgs with a pair of fermions. We find that the inclusive production cross-section of a doubly charged Higgs with a pair of neutrinos is $\sim 3$ times larger than the production cross-section of only a pair of doubly charged Higgs at a $10$ TeV muon collider. Therefore, we stick to the signal of four muons with missing energy in the muon collider. As a lepton collider has been considered here, the fermions must be a pair of charged or neutral leptons. The generation of the signal process is mentioned below:
\begin{equation*}
    \mu^+ \mu^- \to H^{++} H^{--} x~y, H^{\pm\pm} \to \mu^{\pm}\mu^{\pm}
\end{equation*}
where $\{x, y = \mu^{\pm}, \nu_\ell\} $.

Although the dominant contribution to the signal should have come from the vector boson fusion ($W^{+} W^{-}$) and photon fusion, we found that the total cross-section is significantly less than the individual contribution of the vector boson fusion processes. Due to the presence of a very large number of possible diagrams (involving new physics particles), we have considered an inclusive generation of the signal. Due to the very high CM energy, the two neutrinos or the two muons emitted from the primary vertex would be very close to the beam axis, consequently out of reach for the designed detectors. Therefore, not only the neutrinos, but also the muons along the beamline, would contribute to the missing energy in this process.
The following SM processes can mimic our signal: triple vector boson processes ($Z Z Z, Z W^{+} W^{-}, \gamma\,({\rm offshell}) W^{+} W^{-}, \gamma\,({\rm offshell}) Z Z, \gamma\,({\rm offshell})\,\gamma ({\rm offshell}) Z $). For a four-lepton final state with missing energy, the number of events coming from these processes is negligible due to the small leptonic branching ratios of the $W^\pm$ and $Z$ bosons and the additional suppression arising from  $\alpha_{em}\approx {\cal O}(10^{-2})$. These background events are so tiny that they have hardly any impact on the $d_0$ distributions. Hence, 
we only present the signal distribution of the variable $d_0$ for our benchmark points based on cut-based analysis. 

To generate the signal events for the muon collider, the following set of pre-selection cuts has been used:
\begin{equation}
\tt{\left|\eta(\ell)\right| < 3.0;~~~ p_{T}(\ell) > 10~GeV;~~~ \Delta R_{\ell\ell} \geq 0.2}.
\label{eq:preselection_mu}
\end{equation}
Then the events are passed to $\tt{PYTHIA8}$\cite{Sjostrand:2014zea} for hadronization, and the detector simulation is performed using $\tt{DELPHESv3}$\cite{deFavereau:2013fsa} with the help of the muon collider card~\cite{muoncard} for IMCC.
Along with the generational cuts, the following set of cuts are employed to enhance the signal.
\begin{figure}
    \centering      

    \includegraphics[scale=0.35]{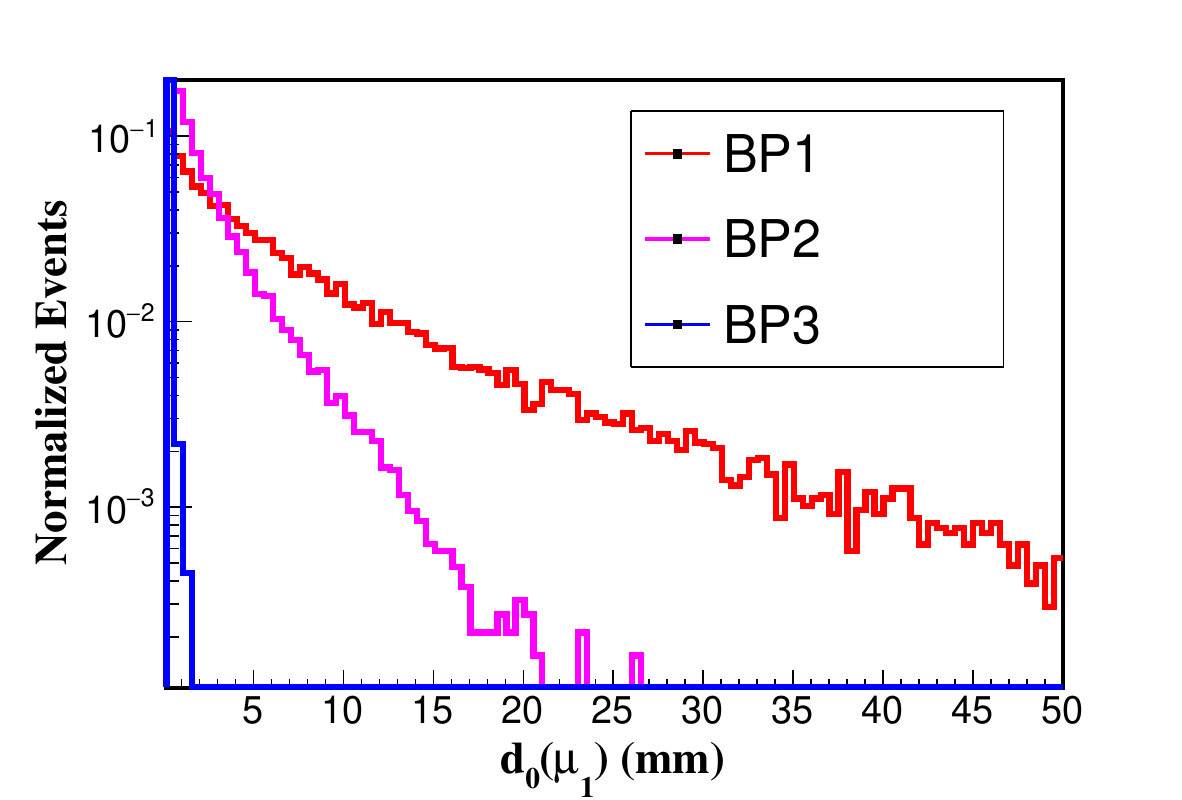}
    \includegraphics[scale=0.35]{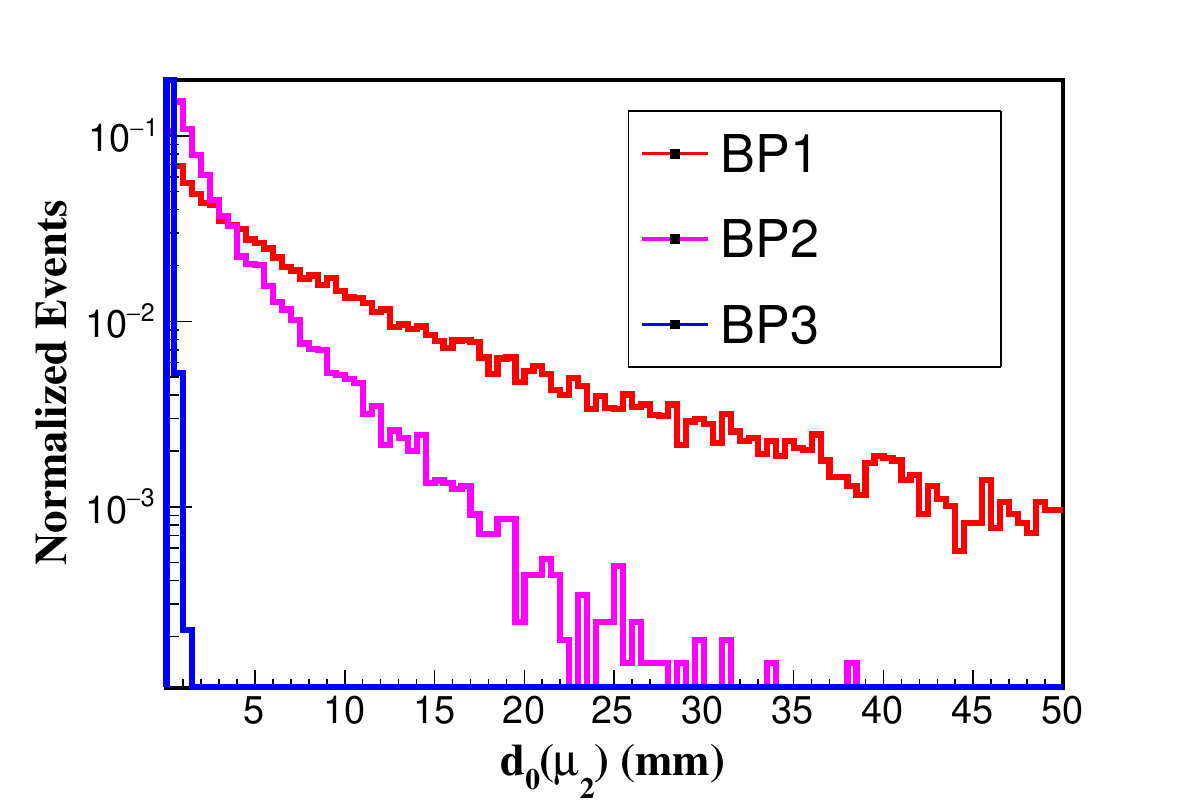}\\
    \includegraphics[scale=0.35]{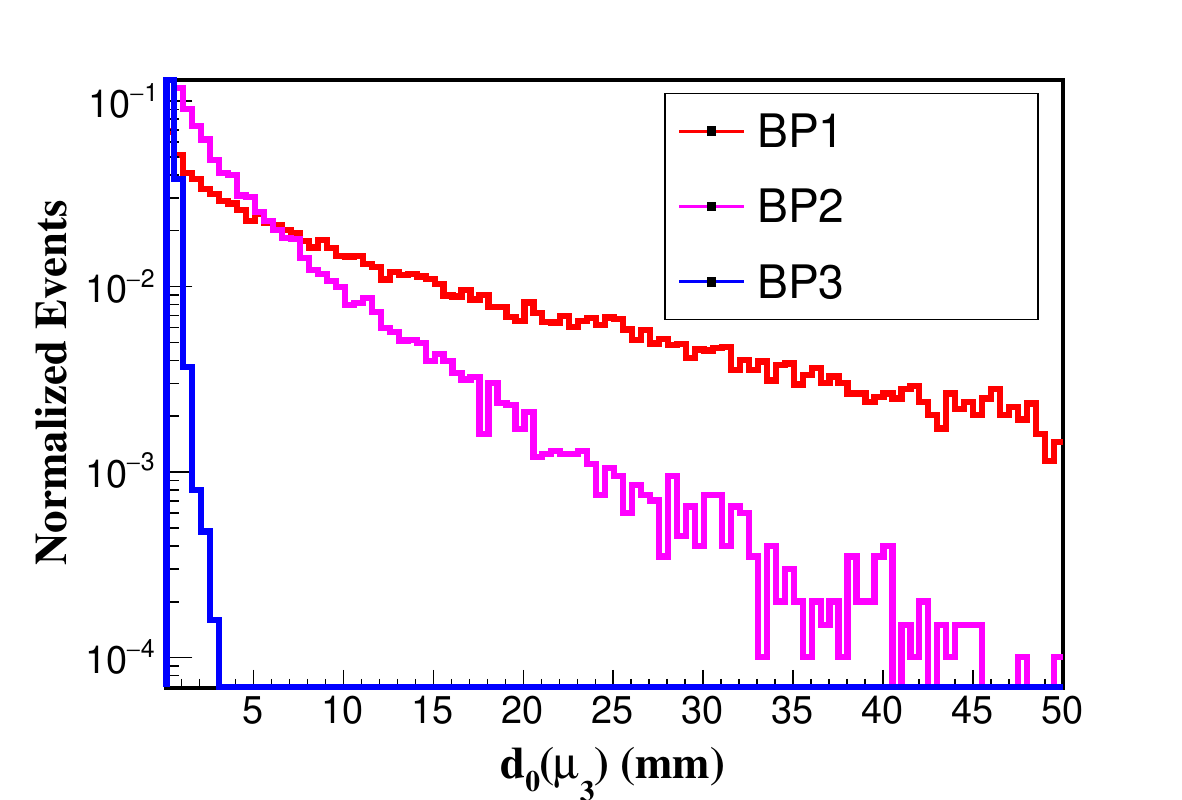}
    \includegraphics[scale=0.35]{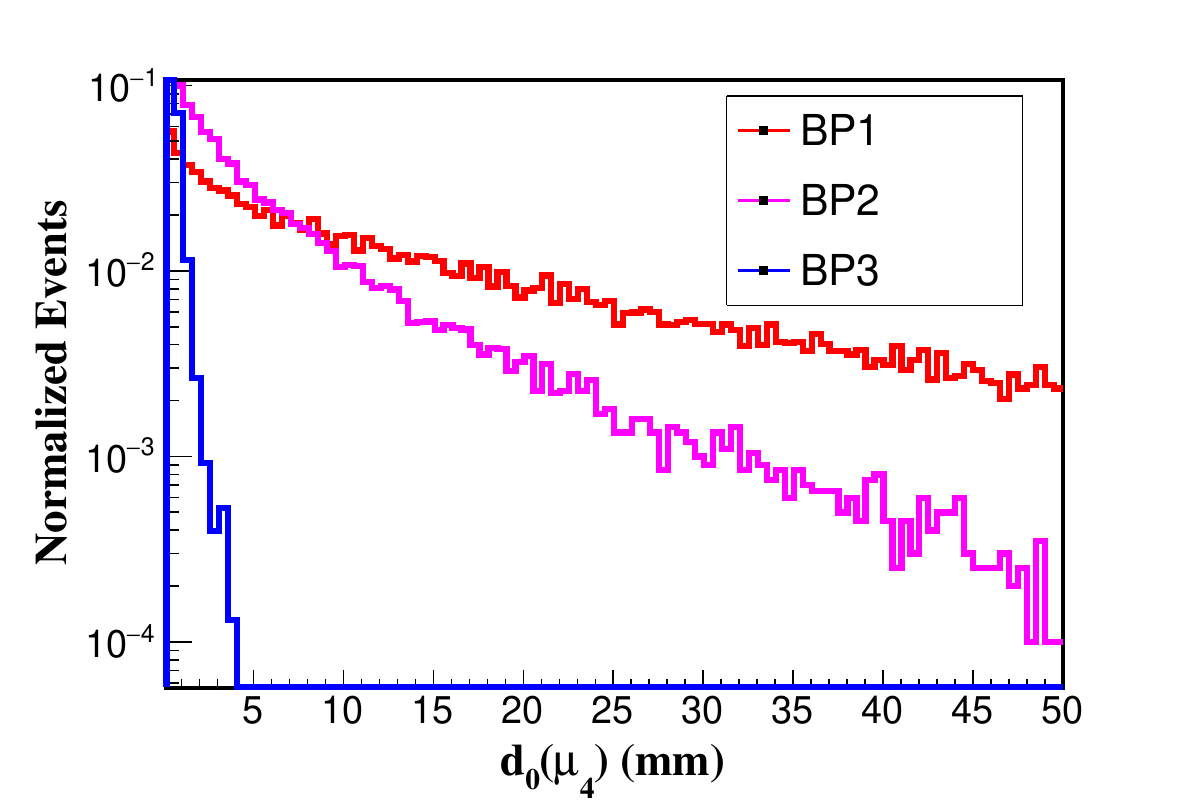}
\caption{Normalized distribution of the $|d_0|$ variable of the leading and sub-leading muons for signal benchmark points BP1, BP2, and BP3 at muon collider with the same CM energy and luminosity as Figure 5.} 
    \label{fig:imcc_d0}
\end{figure} 
\begin{enumerate}
    \item C2-1: In the first step, we select events with exactly four $p_T$ ordered muons.  
    \item C2-2: In the next stage, sequential $p_T$ cuts have been used on muons and events with $p_T(\ell_1)> 30$ GeV, $p_T(\ell_2)> 25$ GeV, $p_T(\ell_3)> 25$ GeV, $p_T(\ell_4)> 20$ GeV have been selected. The corresponding $p_T$ distribution of the muons before applying any cut is presented in Fig.\,\ref{fig:imcc_ptmu}.  
    \item C2-3: The angular coverage of the upcoming muon collider goes from $10^{\circ}$ to $170^{\circ}$ \cite{InternationalMuonCollider:2025sys}, which corresponds to pseudo-rapidity in the range $ -2.46 < \eta < 2.46$. We conservatively select the four muon events with $|\eta (\ell_i)| < 2.2$.
    \item C2-4: Here, we demand a minimum of $\Delta R > 0.2$ cut on the same sign muons. The normalized distribution for the missing energy and missing transverse energy ($\slashed{E}_T$) is depicted in Fig.~\ref{fig:imcc_met}. The charged Higgs pair being very light in comparison to the CM energy, most of the beam energy is deposited in the missing energy, as shown. To avoid fake missing transverse energy events due to the detector resolution effect, we impose a moderate cut on missing transverse energy $\slashed{E}_T>$ 30.0 GeV. 
    \item C2-5: Finally, to investigate the long-livedness of the doubly charged scalar, we put a cut over the $d_0$ variable of the muons while selecting the events. As shown in Fig.~\ref{fig:imcc_d0}, the $|d_0|$ distribution of the benchmark 3 peaks is mainly around the zero value, which is also consistent with the decay length for that particular value of the doubly charged mass. According to the recent proposal of the muon collider detector design \cite{InternationalMuonCollider:2025sys}, as $d_0 < 0.1$ mm is taken for prompt decays, $d_0 > 0.1$ mm can be considered as a signature of long-lived decays. Therefore, the cuts used on the variable $d_0$ of the four muons are, respectively, $|d_0|>$  0.1, 1 mm. The $|d_0|$ distributions of the muons are shown in Fig. \ref{fig:imcc_d0}. 
\end{enumerate}

\begin{figure}
    \centering        
    \includegraphics[scale=0.5]{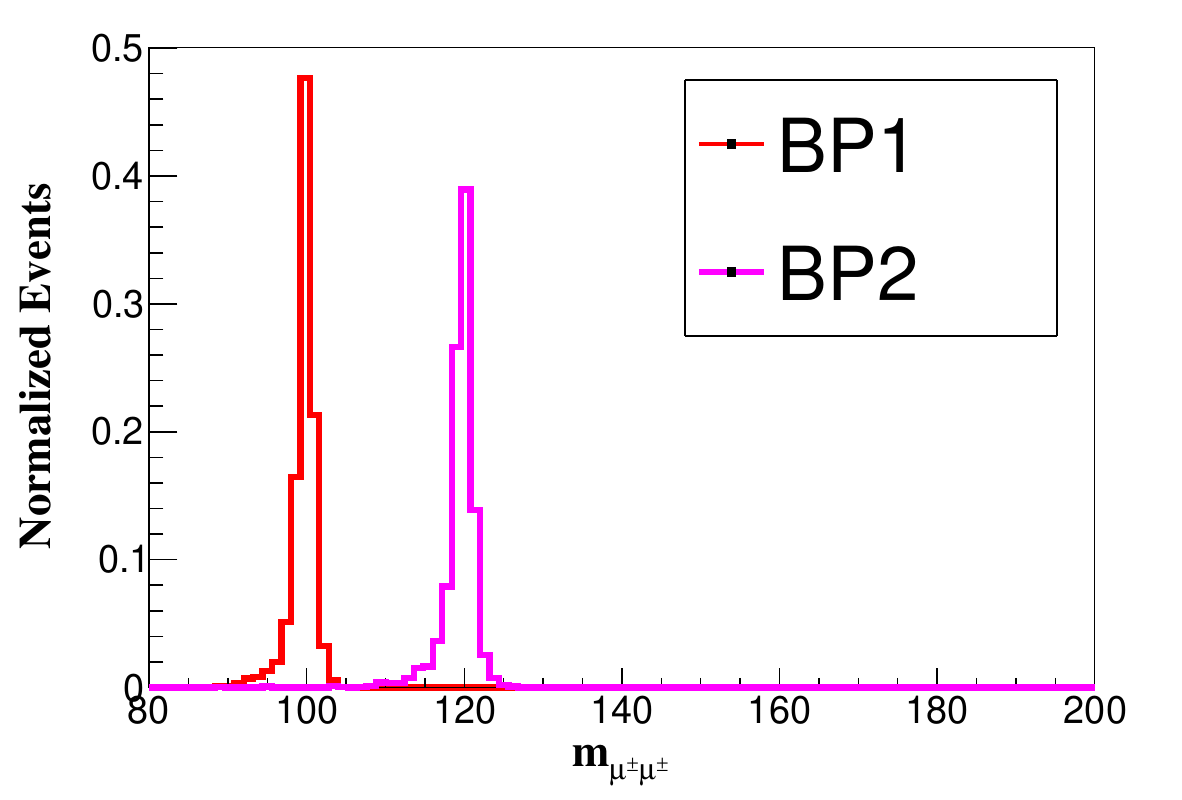}
\caption{Normalized distribution of invariant mass of the same-sign di-muons for BP1 and BP2 at muon collider with the same CM energy and luminosity as Figure 5.} 
    \label{fig:imcc_invar}
\end{figure}

The corresponding SM backgrounds for this process are already very small; as a result, no background events survive after applying the above set of cuts in our simulation. In Tab.~\ref{tab:cutflow_imcc}, we show the number of 
events that survive after giving all the aforementioned cuts at an integrated luminosity of 
10$\rm ab^{-1}$. To show the credibility of the $d_0$ cut, a comparison between $d_0$ and $\rho$ cut is shown in Table~\ref{tab:comp_muon_col}. The integrated luminosity required to see 5 signal events is $\sim 155 (256){\rm fb}^{-1}$
for BP1(BP2), as mentioned in Tab.~\ref{tab:5ev_imcc} using the $d_0$ cut. These results show that a high-energy muon collider 
can also act as a discovery machine for doubly charged scalars, comparable to the ILC in the mass
range (100-150) GeV. In Fig.\ref{fig:imcc_invar}, we show the invariant mass distribution of 
same-sign dimuons for two benchmark points after applying all the mentioned cuts, providing a distinct way for the discovery of a 
doubly charged scalar. 

\begin{table}[]
    \centering
    \begin{tabular}{|c|c|c|c|c|c|c|c|} \hline
      Benchmark points & C2-1 & C2-2 & C2-3 & C2-4 & \multicolumn{2}{c|}{C2-5}  \\ \hline
     & & & & & $d_0 >$ 0.1 mm & $d_0 >$ 1 mm \\ \hline
      BP1 & 512 & 406 & 353 & 343 & 323 & 234 \\ \hline
       BP2  & 338 & 289 & 247 & 239 & 195 & 66 \\ \hline
       BP3 & 36 & 33 & 28 & 27 & $<$1 & $<$1 \\ \hline    
    \end{tabular}
    \caption{Cut flow table showing the number of signal events
    at the muon collider at ${\cal L}_{\rm int} = 10~{\rm ab}^{-1}$ and $\sqrt{s} = 10$ TeV. The last two columns show the final signal events after two choices of $d_0$ cut. }
    \label{tab:cutflow_imcc}
\end{table}

\begin{table}[htbp]
\centering
\begin{tabular}{|c|c|c|c|c|}
\hline
\multirow{2}{*}{Benchmark Point} &
\multicolumn{2}{c|}{$>0.1~\mathrm{mm}$} &
\multicolumn{2}{c|}{$>1~\mathrm{mm}$} \\
\cline{2-5}
& $\rho$ & $d_0$ & $\rho$ & $d_0$ \\
\hline
BP1 & 316 & 323 & 296 & 234 \\
\hline
BP2 & 231 & 195 & 181 & 66 \\
\hline
BP3 & 10 & 0 & 0 & 0 \\
\hline
\end{tabular}
\caption{Comparison of the number of signal events at the muon collider after applying the $d_0$ and $\rho$ cuts.}
\label{tab:comp_muon_col}
\end{table}
    \begin{table}
    \centering
    \begin{tabular}
        {|c|c|} \hline
       Benchmark Points & ${\cal L}_{\rm int}$ required to see 5 events \\
       & after $|d_0| >0.1$mm cut (fb$^{-1}$)\\ \hline
        BP1 & 155 \\ \hline
        BP2 & 256 \\ \hline
        BP3 & - \\ \hline
    \end{tabular}
    \caption{Integrated luminosity required to see 5 events at the 10 TeV muon collider.
    The symbol ``-" for the BP3 benchmark point represents almost zero signal events after all cuts. }
    \label{tab:5ev_imcc}
\end{table}

Here, we comment on the beam-induced background  (BIB) \cite{Bartosik_2020,MOKHOV20122015}, which is a major issue in the muon collider environment. The initial muons decay on their way to flight, producing fluxes of electrons, photons, and neutrinos that create a large background near the beamline. However, here, restricting ourselves to the central region of the detector $(|\eta|  < 2.2)$ and demanding events with only muons, we expect to avoid these background contaminations.

Before concluding this section, we emphasize that a long-lived doubly charged Higgs boson with mass in the range $m_{H^{\pm\pm}} \in [100,160]~\text{GeV}$ can be efficiently probed at both the ILC and IMCC. For heavier $H^{\pm\pm}$ masses, however, the ILC offers superior sensitivity owing to its larger production cross section and finer pixel resolution.

\section{Conclusion}~\label{conc}

In this paper, we investigate the discovery prospects of long-lived doubly charged scalars at future lepton colliders within the framework of the Type-II seesaw model. Focusing on the parameter region where the doubly charged scalar behaves as a long-lived particle, we identified viable benchmark choices that satisfy all current theoretical and experimental constraints and are accessible at the ILC and a high-energy muon collider.
We demonstrate that the search strategies at these facilities are chosen on the basis of the proposed CM of energy of the upcoming colliders. At ILC, with center-of-mass energies of 500~GeV, pair production of doubly charged scalars via the Drell-Yan process provides a clean and efficient probe in the same-sign dimuon channel. At the muon collider operating at 10 TeV energies, the production of $H^{\pm\pm}$ through vector boson fusion becomes dominant, allowing searches in the $\mu^\pm \mu^\pm + \slashed{E}_T$ final state. In both cases, Standard Model backgrounds can be effectively suppressed by exploiting displaced vertex information through a cut on the impact parameter $d_0$. Our analysis shows that long-lived doubly charged scalars with masses in the range $[100$--$180]~\text{GeV}$ can be probed at these future lepton colliders.  We observe that an integrated luminosity of $\sim 3~\mathrm{fb}^{-1}$ is
sufficient to obtain five signal events at the ILC, while $\sim 250~\mathrm{fb}^{-1}$
is required at the muon collider for BP1 and BP2. Since for the choice of benchmark point BP3, the doubly-charged scalar has a very low proper decay length, it cannot be probed at the muon collider and requires integrated luminosity $\cal{O}$ (1) $\rm ab^{-1}$ at the ILC to see at least 5 events.

Furthermore, the clean experimental environment of lepton colliders enables efficient mass reconstruction from same-sign dimuon pairs, significantly enhancing the discovery potential. These results highlight the unique capability of future lepton colliders to explore long-lived particle signatures and provide a compelling physics case for their role in probing extended scalar sectors beyond the SM.

\section*{Acknowledgement}

ND thanks Debajyoti Choudhury, Rameswar Sahu, Purnath Unnikrishnan, Yashasvi and Vineet Kumar Jha for useful physics discussions. ND thanks Amit Adhikary regarding numerical simulations. ND's work is supported by ANRF CRG/2023/008234. NG's work is supported by the Japan Society for the Promotion of Science (JSPS) as part of the JSPS Postdoctoral Program (Standard), grant number: JP24KF0189, and by the World Premier International Research Center Initiative (WPI), MEXT, Japan (Kavli IPMU). NG and ND would also like to thank the Indian Association for the Cultivation of Science(IACS) for the hospitality where part of the project was done.

\appendix
\section*{Appendix}
\section{Partial decay width of the doubly charged scalar}\label{app:partial decay length}

\begin{figure}
    \centering
    \includegraphics[width=0.7\linewidth]{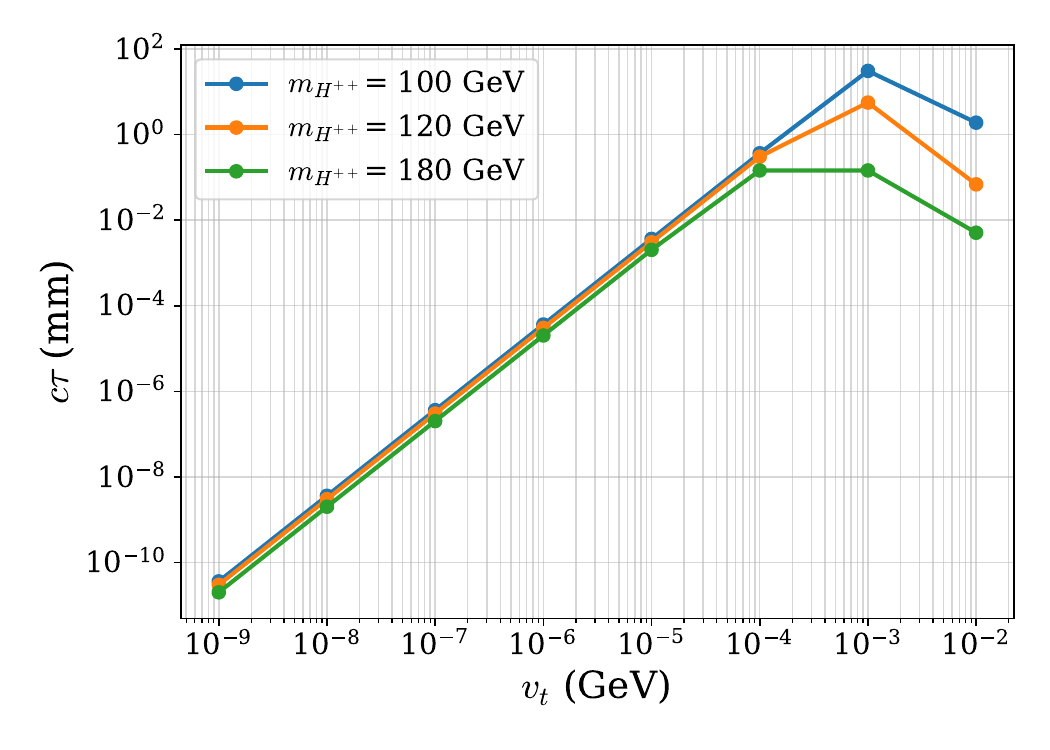}
    \caption{Proper decay length of $H^{++}$ as a function of $v_t$ for our benchmark masses $m_{H^{++}}= 100, 120, 180$ GeV.}
    \label{fig:ctau}
\end{figure}

\begin{figure}
    \centering
    \includegraphics[width=01.0\linewidth]{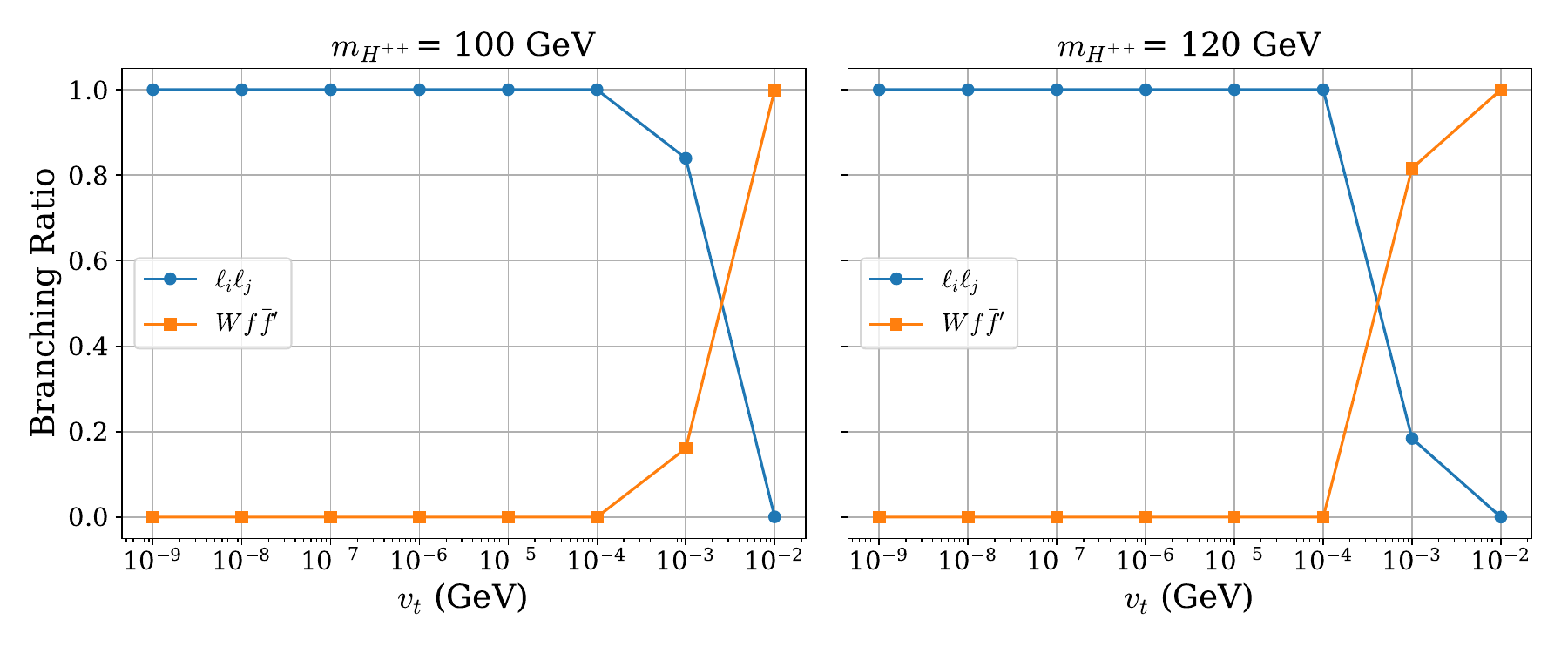}
    \caption{Branching ratio of the doubly charged Higgs in different decay modes}
    \label{fig:BRratios}
\end{figure}

The doubly charged scalar depending on its mass and triplet vev, can decay to a pair of leptons or a pair of $W$ bosons. For a mass region where $m_{H^{\pm\pm}}< 2 m_{W^{\pm}}$ and the partial decay width to leptons is small, $H^{\pm\pm}$ can have cascade decays. The corresponding partial decay widths are given by \cite{BhupalDev:2018tox,Antusch:2018svb}
\begin{equation*}
    \Gamma (H^{\pm \pm} \to l_i^{\pm} l_j^{\pm}) = S \frac{m_{H^{\pm \pm}}}{8 \pi} \frac{|m_{ij}|^2
}{2 v^2_t}\end{equation*}
where S=1(2) when $i=j$ ($i \neq j$) and $m_{ij}$ are neutrino mass matrix elements.
\begin{equation*}
    \Gamma (H^{\pm \pm} \to W^{\pm} (W^{\pm})^* \to W^{\pm} f \bar{f^\prime} ) = \frac{g^6 v^2_t m_{H^{\pm \pm}}}{6144 \pi^3} (3 + N_c \sum_{q , q^\prime} |V_{q, q^\prime}|^2 ) F\Big(\frac{m^2_{W}}{m^2_{H^{\pm \pm}}}\Big)
\end{equation*}
\begin{equation*}
    \Gamma (H^{\pm \pm} \to W^{\pm} W^{\pm}) = \frac{g^4 v^2_t m^3_{H^{\pm \pm}} }{64 \pi m_W^4} \sqrt{1- \frac{4 m^2_{W}}{m^2_{H^{\pm \pm}}}} (1- \frac{4 m^2_{W}}{m^2_{H^{\pm \pm}}} + \frac{12 m^4_{W}}{m^4_{H^{\pm \pm}}})
\end{equation*}
For our chosen benchmark masses,  we have calculated the total decay width of the doubly charged scalar using Madgraph and presented the variation of corresponding $c\tau$s with triplet vev in Fig.\ref{fig:ctau}.  
Here while the rise in the decay length upto $v_t \sim 10^{-4}$GeV is due to the decreasing di-lepton partial decay width, around $v_t \sim 10^{-3}$GeV, the three body cascade decay starts dominating for the first two benchmark masses and the decay length falls accordingly. However, for the third bench mark mass (180 GeV), as production  of a pair of $W$s is kinematically allowed, for larger $v_t$, instead of cascade decays, it takes over the di-leptonic decays. A plot showing the branching ratios of the di-leptonic and cascade decays of $H^{++}$ is shown in Fig.\ref{fig:BRratios}.
Here the choice of lowest mass of the doubly charged scalar is motivated by the LEP constraint on the singly charged scalar mass. With the mass crossing the threshold of $W$ pair production, the decay-length decreases considerably. Therefore the available decay length for the allowed parameter space is roughly $[0.1\sim 30]$mm.     
\section{$\Delta$R distribution of leptons}
\label{app:delta R distribution}
\begin{figure}
    \centering
\includegraphics[scale=0.4]{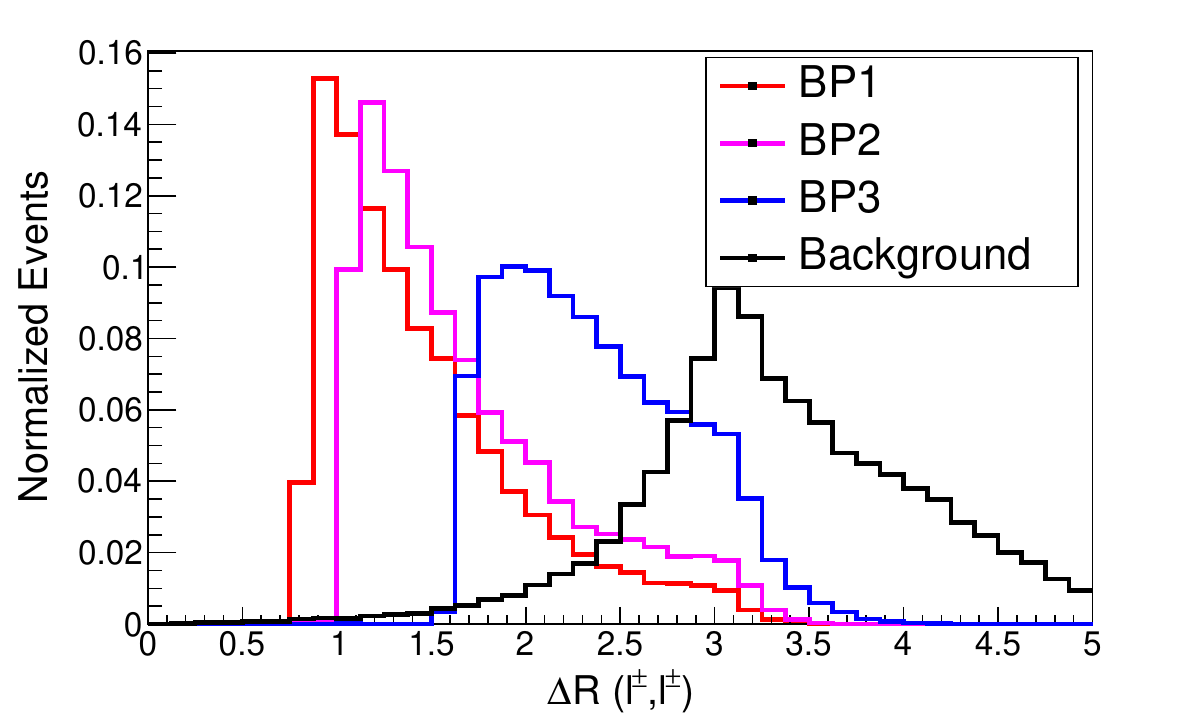}~
\includegraphics[scale=0.42]{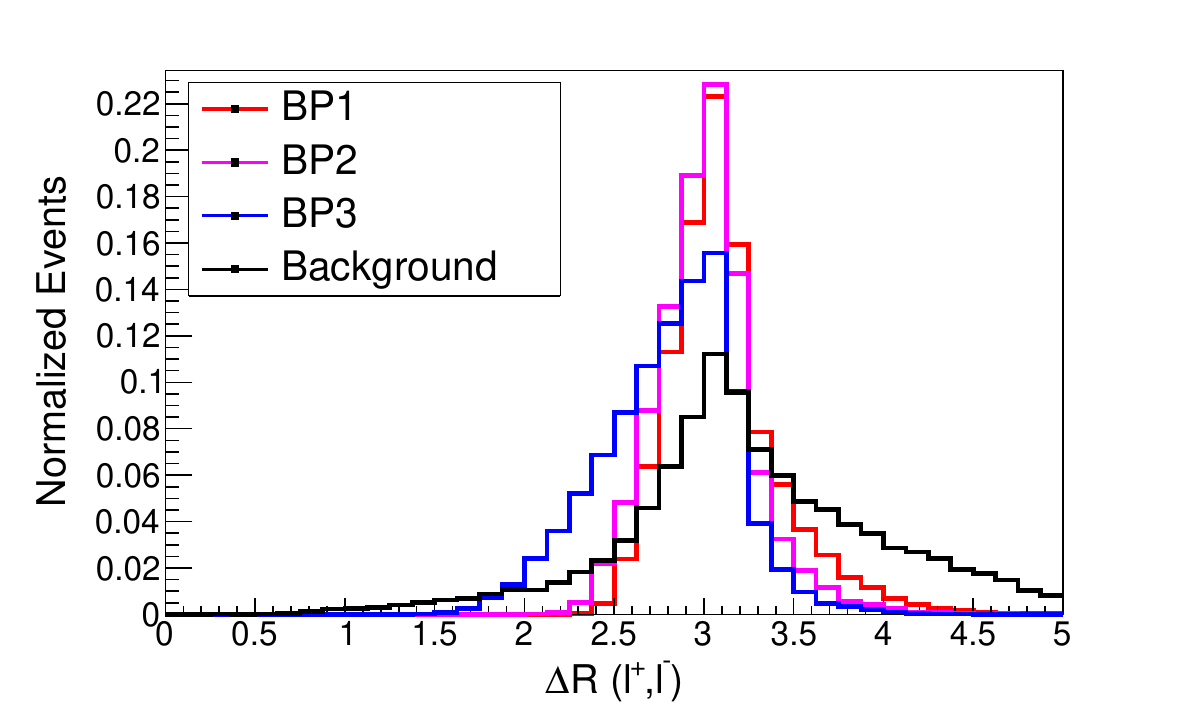}\\
\caption{Normalized distribution of the $\Delta R$ of the same-sign muons (left panel) and opposite sign muons (right panel) for three benchmark points at ILC.}
    \label{fig:deltaR}
\end{figure}

 The $\Delta$R$_{\ell\ell}$ distribution is shown in Fig.~\ref{fig:deltaR}. For the signal, the separation between same-sign leptons peak around 1.0 and for the background it peaks at around $\pi$. For the opposite-sign leptons, the $\Delta$R distribution peaks near $\pi$ both for the signal and the background. We have also explicitly checked that imposing $\Delta R_{\ell\ell}>0.1$ leaves the resulting distributions and final results essentially unchanged. This confirms that the $\Delta R$ requirement adopted in the pre-selection does not have a significant impact on the analysis. 


\providecommand{\href}[2]{#2}\begingroup\raggedright\endgroup

\end{document}